\documentclass[%
 reprint,
 amsmath,amssymb,
 aps,
prx,   
]{revtex4-2}

\usepackage{graphicx}
\usepackage{dcolumn}
\usepackage{bm}
\usepackage{amsthm}
\usepackage{float}
\usepackage{placeins}

\usepackage{adjustbox} 

\usepackage{hyperref}
\usepackage{braket}
\usepackage{svg}
\usepackage[caption=false]{subfig}
\usepackage{amsmath}
\usepackage{amsfonts}
\usepackage{bbm}

\usepackage{xcolor}

\usepackage{tikz}

\begin{document}
\preprint{APS/123-QED}

\title{Variational Learning with Sparse Long-range Entangling Gates}
\author{Helene M. L\"osl}
\author{Aydin Deger}
\author{Andrew J. Daley}
\affiliation{Department of Physics, Clarendon Laboratory, University of Oxford,
Parks Road, Oxford OX1 3PU, United Kingdom}

\date{\today}

\begin{abstract}
The performance of variational quantum algorithms depends in general on the structure of the parametrized quantum circuit, but the most common ans\"atze are typically based on local couplings. Motivated by the extended connectivity available with neutral atoms and trapped ions, we examine when structured long-range connectivity provides a useful resource, focusing on sparse power-of-two (PWR2) coupling graphs. Using dynamical Lie-algebra analysis, approximate unitary-design diagnostics, and finite-depth measures of expressibility and entanglement, we examine how these geometries enlarge the accessible operator space. This enlarged space alone is not sufficient to ensure trainability of the parameterized circuit for given target problems, and we explore performance across example problems with and without long-range coupling, identifying where sparse coupling graphs are or are not likely to provide an advantage.
We also introduce a variational scheme that maps hierarchical long-range Hamiltonians to geometrically local ones that can be optimized with short-range circuits. Together, these results identify circuit geometry and qubit reconfigurability as task-dependent resources for variational algorithms, relevant to ongoing developments in quantum hardware with long-range connectivity.
\end{abstract}
\maketitle

\section{\label{sec:level1}Introduction}
Variational quantum algorithms (VQAs) constitute one of the central paradigms for near-term quantum computation, with prominent applications in ground-state preparation and simulating non-equilibrium many-body dynamics \cite{yuan_theory_2019,kokail_self-verifying_2019,cerezo_variational_2021}. Their performance critically depends on the choice of a parametrized quantum circuit (PQC). The PQC must be expressive enough to represent physically relevant states, trainable by classical optimization, and compatible with hardware. Balancing these requirements when designing circuit architectures remains a major challenge.

Many widely used variational ansätze employ a combination of single-qubit and nearest-neighbor two-qubit gates \cite{defenu_long-range_2023,lyu_variational_2023}. This design provides a straight-forward structure to analyze, and aligns with the local connectivity on those hardware platforms where long-range operations incur high overhead \cite{baumer_measurement-based_2025,linke_experimental_2017}. However, while this choice has advanced understanding of trainability, expressibility, and the barren-plateau phenomenon \cite{mcclean_barren_2018,cerezo_variational_2021}, it is interesting to ask whether the resources to express states of longer-range correlations and entanglement could be more naturally expressed if we utilize the possibility of plaforms such as trapped ions and neutral-atom arrays, which naturally support high-fidelity entangling gates between distant qubits. These use collective motional modes \cite{monroe_programmable_2021, foss-feig_progress_2025,ransford_98-qubit_2026} or atom tweezer reconfigurability and/or long-range Rydberg interactions \cite{bluvstein_logical_2024,finkelstein_universal_2024,muniz_high-fidelity_2025}. This is a natural complement to the role that long-range interactions play in many atomic physics systems, which has motivated much recent work on related models \cite{defenu_long-range_2023}.

In this work, we build on these observations to explore sparse long-range entangling ansatz geometries as a variational resource. Rather than focusing on fully connected or randomly wired circuits \cite{lyu_variational_2023}, we study sparse, deterministic long-range architectures defined on a power-of-two (PWR2) coupling graph \cite{hung_quantum_2016,bentsen_fast_2019,qin_characterizing_2019,periwal_programmable_2021,mivehvar_cavity_2021}. The PWR2 graph, like a hypercube, sits between nearest-neighbor (NN) and all-to-all (A2A) connectivity. It supports fast information scrambling, in that it requires only $\log(n)$ steps to connect any two nodes on an $n-$qubit graph, and tunable entanglement growth, while only having $\log(n)$ interaction terms per spin \cite{hashizume_deterministic_2021,hashizume_tunable_2022,kuriyattil_onset_2023}. We note that both all-to-all and sparse interactions have been previously realized in experiments \cite{britton_engineered_2012,islam_emergence_2013,zeiher_coherent_2017,vaidya_tunable-range_2018,hollerith_realizing_2022}.

By leveraging this structure, we demonstrate that PWR2 circuits can, in some cases, offer a favorable balance between expressibility and depth for the variational quantum eigensolver (VQE) on the PWR2 transverse field Ising model (TFIM) and can be used for learning quench dynamics of the NN TFIM. In analyzing this we note that there are two separate questions that need to be addressed. The first is the ability of an ansatz within a small number of layers to represent the correlations and entanglement present in the relevant state. For moderate to large system sizes, an ansatz with long-range connectivity has a natural advantage in being able to express different correlations with a relatively small circuit depth. However, the question of whether the ansatz can be efficiently trained to represent the correlations particular in a chosen scenario is more difficult and problem-dependent. We discuss this in example cases below.

We then also introduce a second route to accessing long-range correlations by optimizing in a permuted space. This approach complements the use of long-range gates in the variational ansatz, as it also relies on architectures capable of dynamically reconfiguring qubits. In quantum many-body systems, qubit permutations can turn long-range interaction patterns into local ones. The Monna map is one example. It reindexes degrees of freedom according to a tree-like geometry \cite{bentsen_treelike_2019, hashizume_deterministic_2021}. This enables variational optimization in a localized representation via short-range circuits. After optimization, we recover long-range-correlated states by applying an exact inverse transformation during state reconstruction and measurement on the quantum device.

Together, these two approaches define a framework for depth-efficient, hardware-aligned variational learning. We draw on insights from circuit geometry, dynamical Lie algebras, and expressibility diagnostics. This shows how nontrivial connectivity can improve learning performance without relying on fully random or all-to-all architectures. Our results identify circuit geometry as a physically motivated and experimentally relevant design axis for variational quantum algorithms.

Using the power-of-two sparse spin model as an example, we show that VQE for long-range interacting spin systems benefits from incorporating the interaction structure of the Hamiltonian into the variational ansatz. In particular, ansätze with connectivity matching the underlying sparse couplings achieve lower energy errors with shorter circuit depths and fewer parameters. Notably, this benefit is not limited to the long-range regime: even in the short-range limit, the long-range ansatz performs well at the small system sizes considered, showing that added connectivity does not reduce performance.

We also examine connectivity by using hardware reconfigurability to make the problem short-ranged. In our framework, we first map the problem to a connectivity-adapted form. Then, we solve the new variational problem classically or quantumly and reconstruct the original state on the quantum device.

To quantitatively assess PWR2 circuits as sparse long-range ansätze, we examine their expressibility and trainability. This includes calculating the connectivity-dependent second-moment superoperator and analyzing entangling power with the Meyer–Wallach entropy. Together, these results support viewing qubit connectivity as both a hardware constraint and a structural feature for designing and analyzing variational learning protocols.

The rest of the paper is structured as follows. In Sec.~\ref {sec:pwr2_circuits}, we define and introduce the PWR2 graph and circuits as the central objects of this paper. Sec.~\ref{sec:theory} then theoretically characterizes the PWR2 circuits in terms of their expressibility and trainability, before Sec.~\ref {sec:vqe_pwr2} and Sec.~\ref{sec:vqa_tfim} consider two applications of this connectivity. In Sec.~\ref{sec:monnamap}, we turn to the Monna map to localize problems on the PWR2 graph. We end with a discussion of the results in Sec.~\ref{sec:discussion}.

\begin{figure*}
    \includegraphics[width=0.95\linewidth]{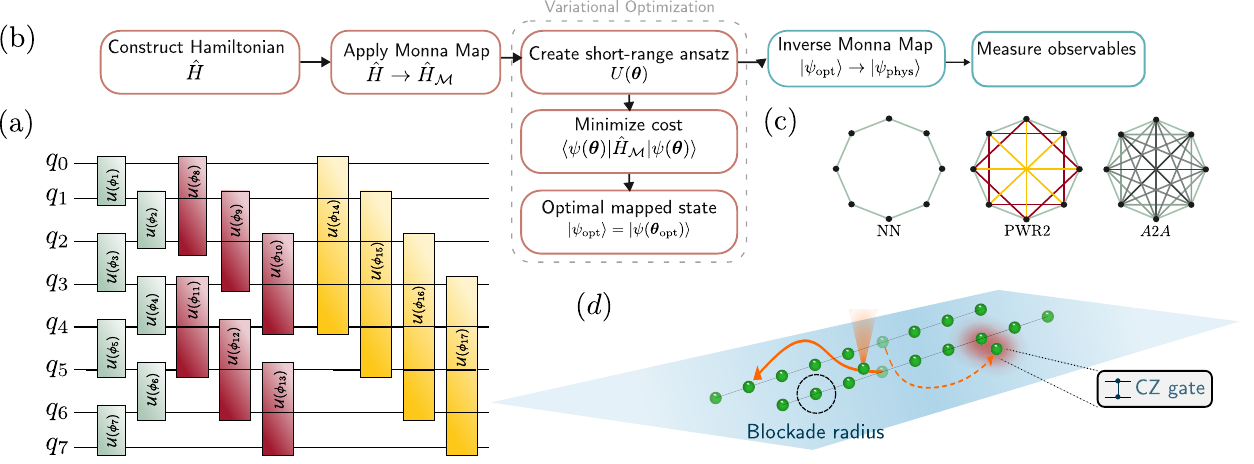}
    \caption{(a) parametrized quantum circuit for eight qubits, different colors represent distances between entangled qubits  $d=2^0=1$ (green),$2^1=2$ (red), $2^2=4$ (yellow) (b) Flowchart showing the Monna variational pipeline that allows solving a variational problem in a localized representation before accessing the solution via inverting the unitary map. Red indicates steps that can be performed classically, and blue indicates steps executed on quantum hardware. (c) Coupling graphs considered in the main text, from left to right, the nearest neighbor (NN), power-of-two (PWR2), and all-to-all (A2A) connected graph. Edges connecting qubits separated by the same distance are shown in the same color. (d) Implementation of a CZ gate between geometrically distant atoms held in a light sheet and shuttled using mobile Rydberg tweezers.}
    \label{fig:overview}
\end{figure*}
\section{Structured long-range variational circuits}
\label{sec:pwr2_circuits}

 The PWR2 coupling graph is defined on
$V=\{1,2,\dots,n\}$, with edge set
\begin{align}
E_{\mathrm{PWR2}}
=
\Bigl\{\{i,\, i+2^k \!\!\!\pmod n\}\;\Big|\;
i=1,\dots,n,\; \\
k=0,\dots,\lfloor \log_2(n/2)\rfloor
\Bigr\},\nonumber
\end{align}
where edges are unordered, and duplicate edges are counted only once. We use periodic boundary conditions and identify qubit labels modulo $n$, so
that $i+n\equiv i$. Thus, the graph contains edges between qubits whose separation in the one-dimensional is a power of two,
including nearest neighbors $d=2^0=1$. The number of distinct interaction
distances grows logarithmically with system size, and the graph has logarithmic
degree, giving $\mathcal{O}(n\log n)$ edges rather than the $\mathcal{O}(n^2)$ edges of an
all-to-all graph. If $n$ is even and $d=n/2$ is included, this distance class
contributes only $n/2$ distinct undirected edges rather than $n$.

The nearest-neighbor ring has the edge set
\begin{equation}
E_{\mathrm{NN}}= \bigl\{\{i,i+1 \!\!\!\pmod n\}\mid i=1,\dots,n\bigr\},
\end{equation}
and is a subgraph of $E_{\mathrm{PWR2}}$.

From a physical perspective, the coupling graph plays a role analogous to an interaction geometry, shaping how correlations and entanglement propagate under unitary dynamics. Spin models and Clifford circuits defined on PWR2 graphs have been shown to exhibit several features normally associated with much more dense connectivity, including fast information spreading and long-range entanglement structure~\cite{bentsen_treelike_2019}. In the variational setting, this motivates studying how sparse long-range connectivity modifies the space of states and unitaries accessible at finite circuit depth.

For parametrized circuits, the coupling graph specifies only the available two-qubit supports; it does not by itself define a circuit depth or gate ordering. To construct a layered ansatz, we decompose the PWR2 edge set into periodic distance classes
\begin{equation}
E_d
=\bigl\{\{i,i+d \!\!\!\pmod n\}\mid i=1,\dots,n\bigr\},
\end{equation}
for $d\in\{1,2,4,\dots\},\, d\le n/2$.
 For each distance $d$, the corresponding edges are partitioned into matchings, or sets of disjoint gates, that can be applied in parallel. In circuit diagrams, edge colors denote the distance class $d$,
while the horizontal order of colored layers denotes the chosen gate ordering.

\subsection{Structural Characterization\label{sec:theory}} 

In general, it has proven difficult to predict the performance of variational circuits and the hardness of learning tasks. The two main factors contributing to this challenge are expressibility and trainability \cite{sim_expressibility_2019,cerezo_cost_2021, holmes_connecting_2022}. Expressibility is a circuit's ability to represent various quantum states, determining the size of the state space \cite{sim_expressibility_2019, holmes_connecting_2022}. Trainability, in contrast, refers to how easily the circuit can be optimized using classical methods \cite{cerezo_cost_2021}. These properties are linked, with a known trade-off: higher expressibility broadens the state space but often reduces trainability, as the cost function value changes little across the parameter space, producing a flatter loss landscape. Such flatness can also result from the choice of initial states \cite{mcclean_barren_2018,abbas_power_2021} or cost function \cite{cerezo_cost_2021,larocca_diagnosing_2022}.

\subsubsection{Asymptotic Controllability}
Let us first consider how increased expressibility, caused by changes in qubit connectivity, affects circuit behavior, by identifying the dynamical Lie algebra (DLA) generated by the elementary
parametrized evolutions of the ansatz. This determines the set of unitaries that
can be reached asymptotically, i.e., in the limit of sufficiently many layers and
arbitrary parameters.

Given traceless Hermitian generators $\{H_\mu\}$, corresponding either to
controllable Hamiltonian terms or to parametrized gates $e^{-i\theta_\mu H_\mu}$, we define the dynamical Lie algebra as the real Lie algebra
\begin{equation}
\mathfrak g
=
\operatorname{Lie}_{\mathbb R}\langle -iH_\mu\rangle
\subseteq \mathfrak{su}(2^n).
\end{equation}
Equivalently, $\mathfrak g$ is the smallest real Lie algebra containing the
skew-Hermitian generators of the available elementary evolutions.

For the transverse-field Ising model (TFIM) with generator set
$\{Z_iZ_j:\{i,j\}\in E\}\cup\{X_i:i=1,\dots,n\}$, the classification results of Refs.~\cite{wiersema_classification_2024,kokcu_classification_2026} imply that, under the graph assumptions of Theorem III.9 in Ref.~\cite{kokcu_classification_2026}, the PWR2 graph has the same DLA as the complete graph. For periodic boundary conditions, this gives
\begin{equation}
\mathfrak g_{\mathrm{NN}} \cong \mathfrak{so}(2n)^{\oplus 2},
\qquad
\mathfrak g_{\mathrm{PWR2}} \cong \mathfrak g_{\mathrm{A2A}}
\cong \mathfrak{su}(2^{n-1})^{\oplus2}.
\end{equation}
The direct-sum structure reflects the decomposition into fermion-parity sectors.
Thus, the nearest-neighbor TFIM generators exhibit only quadratic DLA growth, reflecting the free-fermion structure of the model. In contrast, the PWR2 and all-to-all generator sets have exponentially growing DLAs.

While larger DLAs are associated with increased expressibility, this benefit is accompanied by a well-known tradeoff: in sufficiently deep circuits, increased expressibility can lead to regions with exponentially suppressed gradient variance in the cost landscape, a phenomenon known as barren plateaus~\cite{larocca_diagnosing_2022,holmes_connecting_2022}.  From this perspective, one might expect that the larger DLA of the PWR2 model limits the trainability compared to the NN model.
However, two qualifications limit this conclusion. First, the exponential separation between the algebra sizes of the NN and PWR2 Ising models arises from the integrable structure of the NN TFIM and is not a generic consequence of the PWR2 connectivity. Second, DLA-based barren-plateau results typically apply in the regime where the circuit ensemble forms a unitary 2-design on the relevant dynamical Lie group, which may not hold in the finite-depth regime considered in our numerical experiments \cite{holmes_connecting_2022}.

We therefore separate two questions: what is the full asymptotic DLA generated by the ansatz, and what is the commutator depth required to generate the target algebra relevant for a given learning task. We look at several examples to consider practical trainability later in section \ref{sec:applications}. 

As an initial comment, though we can look at the restricted case where the target unitary is generated by the nearest-neighbor TFIM. Then the target dynamics lie inside the quadratic Majorana algebra associated with the NN model, $\mathfrak{so}(2n)^{\oplus 2}$. The question is whether adding gates from the same fate family but with sparse long-range connectivity to the generator set can help to span the target DLA faster or shorter commutator depth, respectively.

For an NN Hamiltonian variational ansatz (HVA) with generators
\begin{equation}
\{Z_iZ_{i+1}:i=1,\dots,n\}\cup\{X_i:i=1,\dots,n\},
\end{equation}
where $i+1$ is understood modulo $n$, the minimal commutator depth required to
generate the quadratic Majorana algebra scales as
\begin{equation}
q=\mathcal O(n).
\end{equation}
More precisely, define the nested commutator spaces
\begin{equation}
\mathcal{L}_0 := \operatorname{span}_{\mathbb R}\{\, -iH_\mu \,\},
\qquad
\mathcal{L}_{k+1}:=\mathcal{L}_k+[\mathcal{L}_k,\mathcal{L}_0].
\end{equation}
The smallest $q$ such that $\mathcal{L}_q=\mathfrak g_{\mathrm{target}}$ is the
commutator depth required to generate the target algebra.

By contrast, if the available Majorana bilinears are supported on the PWR2 graph, Appendix~\ref{app:A} shows that the corresponding commutator depth for the quadratic Majorana target algebra satisfies
\begin{equation}
q=\mathcal O(\log n).
\end{equation}
It shows
that the $\mathfrak{so}(2n)$ algebra relevant for NN TFIM dynamics is generated
from the larger generator set at smaller commutator depth. In this algebraic sense,
PWR2 connectivity reaches the relevant target operators more rapidly than NN connectivity. 

\subsubsection{Approximate unitary 2-design diagnostics}
The DLA characterizes asymptotic reachability, but it does not by itself describe the behavior of finite-depth circuits \cite{zeier_symmetry_2011,zimboras_symmetry_2015,ragone_lie_2024}. To obtain a complementary diagnostic, we study the rate at which circuit ensembles with a specified connectivity approach an approximate unitary 2-design . This calculation is not intended to reproduce the exact variational circuits used in the numerical applications below. Instead, we use random two-qubit gates on the same coupling graph as an idealized probe of the effect of connectivity alone. In this way, differences between NN and PWR2 circuits arise only from the graph geometry and gate ordering, not from a
particular choice of parametrized gate family.

For a coupling graph $G$, we first decompose the edge set into an ordered sequence of matchings,
\begin{equation}
    E(G)=M_1\cup M_2\cup\cdots\cup M_r,
\end{equation}

\noindent where each matching $M_a$ consists of disjoint edges and can therefore be applied
in parallel. A single random block is defined as
\begin{equation}
    U_{\mathrm{block}}(G)=\prod_{a=1}^{r}\prod_{(i,j)\in M_a}U_{ij}^{(a)},
\end{equation}

\noindent where the gates $U_{ij}^{(a)}$ are independently drawn from the Haar measure on $SU(4)$ and embedded on qubits $i$ and $j$. For the PWR2 graph, the matchings are organized by distance classes $d=1,2,4,\dots, \log_2(n/2)$, and different orderings of these distance classes define different PWR2 block orderings.

The corresponding second-moment operator is
\begin{equation}
   M_G^{(2)}=\mathbb{E}_{U_{\mathrm{block}}(G)}\left[U_{\mathrm{block}}^{\otimes 2}\otimes\left(U_{\mathrm{block}}^\dagger\right)^{\otimes 2}\right]. 
\end{equation}

\noindent We compare this operator with the Haar second moment on the full $n$-qubit
Hilbert space,
\begin{equation}
   A_G^{(2)} = M_G^{(2)}-M_{\mathrm{Haar}}^{(2)}. 
\end{equation}
The relevant quantity is the largest singular value of $A_G^{(2)}$ on the nontrivial subspace,
\begin{equation}
    \lambda_G=\left\|A_G^{(2)}\right\|_{\mathrm{nt}},
\end{equation}
where the subscript indicates that the Haar-invariant sector is removed. A smaller value $\lambda_G$ implies a faster contraction of second moments under repeated application of the random block.

Using the result of Ref.~\cite{ragone_lie_2024}, the number of repeated blocks needed to form an $\varepsilon$-approximate unitary 2-design is bounded by
\begin{equation}
   L\geq\frac{\log(1/\varepsilon)}{\log(1/\lambda_G)}. 
\end{equation}

\noindent This expression should be interpreted as an upper-bound diagnostic for the approach to design-like behavior. It does not directly predict optimization performance, nor does it imply that the parametrized circuits used below form 2-designs at the plotted depths.

We evaluate $\lambda_G$ numerically for the NN and PWR2 graph-local ensembles. The moment operators are constructed from exact two-qubit Haar twirls and multiplied according to the matching decomposition of each block; no Monte Carlo sampling over random gates is used. Details of the construction and of the projection onto the nontrivial subspace are given in Appendix~\ref{app:weingarten}. Figure~\ref{fig:entangling_Cap_combined} shows the resulting bound on $L$ as a function of $\varepsilon$. For the system sizes studied here $n\leq16$, the PWR2 block has a smaller nontrivial second-moment singular value than the NN block, leading to a smaller upper bound on the number of blocks required to approach a unitary 2-design. This supports the interpretation that sparse long-range connectivity accelerates the onset of design-like second-moment behavior in this idealized graph-local ensemble.

\begin{figure}[h!]
    \centering
    \includegraphics[width=\linewidth]{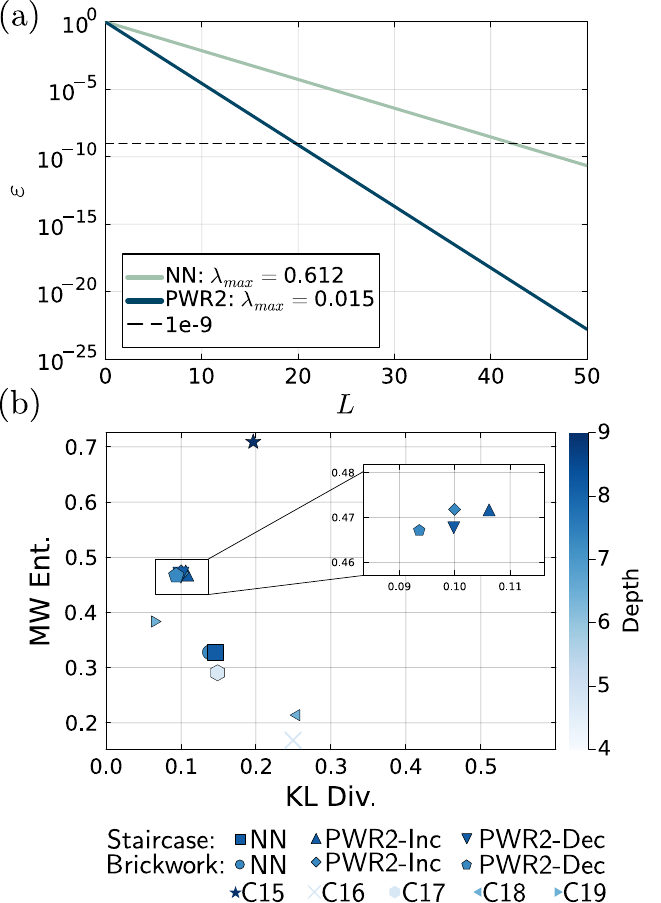}
    \caption{(a)Distance of an ensemble of PQC on $n=16$ qubits from forming an $\varepsilon$-approximate 2-design, or equivalently, the circuit depth $L$ needed to reach a particular distance of a 2-design by evaluating the bound of \cite[Theorem 2]{ragone_lie_2024} for brickwork circuits defined on the NN and the PWR2 coupling graph, respectively. For a given $\varepsilon$, the depths for the PWR2 circuit are substantially lower than for the NN circuit. (b)Expressibility and entangling capability of NN and PWR2 graphs for $n=4$ qubits. Expressibility is quantified through the KL divergence Eq.~\ref{eq:DKL} and entangling capability through the Meyer-Wallach measure Eq.~\ref{eq:MWM}. The color of the markers indicates the depth of the circuit, ranging from 4 to 9 sequential parallel layers, as shown in the colorbar on the right. Circuit ensembles include 5000 independent parameter sets, and fidelity density distributions are computed with 75 bins.}
    \label{fig:entangling_Cap_combined}
\end{figure}

\subsubsection{Expressibility and Entangling capabilities}
To quantify how the algebraic and design-theoretic differences discussed above manifest at finite circuit depth, we employ empirical diagnostics that probe the expressibility and entanglement-generating capability of parametrized quantum circuits. 

We quantify expressibility following the approach introduced in Ref.~\cite{sim_expressibility_2019}, which compares the distribution of fidelities generated by the circuit to that of Haar-random states. Concretely, expressibility is quantified by the Kullback--Leibler divergence
\begin{equation}
    D_{\mathrm{KL}}(P_F \,\|\, P_{\mathrm{Haar}})
    =
    \sum_j P_F(j)\,
    \log\!\left(\frac{P_F(j)}{P_{\mathrm{Haar}}(j)}\right), \label{eq:DKL}
\end{equation}
where $P_F$ denotes the fidelity distribution of the parametrized circuit ensemble and $P_{\mathrm{Haar}}$ is the corresponding distribution for Haar-random states. For the Haar ensemble on an $n$-qubit Hilbert space with dimension $d=2^n$, the probability density of fidelities is analytically  known as\cite{zyczkowski_average_2005} 
\begin{equation}
    P_{\mathrm{Haar}}(F) = (d - 1)(1 - F)^{d - 2}.
\end{equation}
The circuit fidelity distribution $P_F$ is obtained by independently sampling parameter sets $\boldsymbol{\theta}_1$ and $\boldsymbol{\theta}_2$ and evaluating the pairwise fidelities
\begin{equation}
    F
    =
    \big|
    \langle 0 | U^\dagger_{\boldsymbol{\theta}_1} U_{\boldsymbol{\theta}_2} | 0 \rangle
    \big|^2 .
\end{equation}
 Smaller values of $D_{\mathrm{KL}}$ indicate that the circuit ensemble more closely approximates the Haar fidelity distribution and is therefore more expressive in the sense of Ref.~\cite{sim_expressibility_2019}.

To assess the ability of the circuit to generate entanglement, we employ the Meyer--Wallach measure~\cite{meyer_global_2002,brennen_observable_2003}, which measures the single-qubit entanglement. For a pure $n$-qubit state $\rho$, this measure is defined as
\begin{equation}
    E(\rho)
    =\frac{2}{n}
    \sum_{j=1}^n \bigl( 1 - \mathrm{Tr}[\rho_j^2] \bigr), \label{eq:MWM}
\end{equation}
where $\rho_j = \mathrm{Tr}_{\bar{j}}[\rho]$ denotes the reduced density matrix of qubit $j$. The Meyer--Wallach measure takes values in $[0,1]$, with larger values corresponding to higher degrees of entanglement capability.

Figure~\ref{fig:entangling_Cap_combined} compares the numerical results of the two diagnostics for nearest-neighbor and power-of-two circuit geometries at $n=4$ qubits and matched circuit depth, both for the staircase and brickwork ordering of the gates (see App.~\ref{app:ansatze} for a depiction of the circuits). Each ansatz consists of an initial layer of single-qubit rotations $R_y(\alpha_i)R_z(\beta_i)$ applied to all qubits, followed by $L$ layers of entangling CRZ gates determined by the connectivity graph, and a final layer of single-qubit $R_y R_z$ rotations.
For each pair $(i,j)$ with $i<j$, qubit $i$ is taken as the control and $j$ as the target. The circuit depth is defined as the total number of sequential parallel layers, including both single-qubit and two-qubit gate layers, where all gates within a given layer act on disjoint qubits and can therefore be executed simultaneously. All variational parameters are sampled independently from a uniform distribution on $[0,2\pi]$. 

For the PWR2 ansätze, we additionally distinguish between results for increasing and decreasing distances in the gate ordering. We also provide comparisons with some of the example circuits for testing expressibility considered in \cite{sim_expressibility_2019,strobl_qml_2025}. The expressibility and Meyer-Wallach measure are computed using the \texttt{qml-essentials} package \cite{strobl_qml_2025} with $n_\mathrm{samples}=5000$ realizations of each circuit to match the results provided in \cite{strobl_qml_2025}. The empirical fidelity distribution is estimated using a histogram with $n_\mathrm{bins}=75 $ equally spaced bins in $[0,1]$. The Haar distribution is discretized by integrating the analytic density over the same bins. The Meyer–Wallach measure is averaged over the same ensemble of $n_\mathrm{samples} = 5000$ randomly sampled parameter sets.

Across the depths considered, circuits defined on the PWR2 coupling graph exhibit both lower Kullback--Leibler divergence and higher average Meyer--Wallach entropy than their NN counterparts. This quantifies how the sparse long-range connectivity of the PWR2 graph enhances the circuit’s ability to explore larger regions of Hilbert space and to generate entanglement at fixed depth, consistent with the faster design-formation and larger reachable operator space identified in the preceding sections.

At the same time, these results highlight the nuanced relationship between expressibility and trainability. While highly expressive circuits approach Haar-random behavior, such regimes are known to be associated with exponentially suppressed gradient variance and barren plateaus~\cite{cerezo_variational_2021}. The observed advantage of PWR2 circuits therefore arises in an intermediate regime, where expressibility and entanglement are enhanced relative to nearest-neighbor architectures without fully entering the Haar-random limit. At the finite depths studied here, PWR2 circuits generate Haar-like fidelity statistics and entanglement more rapidly than NN circuits. These diagnostics indicate enhanced finite-depth representational capacity. However, neither the dynamical Lie algebra, the approach to a unitary 2-design, nor empirical expressibility directly predict variational optimization performance. Rather, they characterize different structural properties of the circuit family. Whether these properties translate into improved learning depends on the target problem, the cost function, and the optimization landscape. The numerical studies in the following sections therefore test when the enhanced representational capacity associated with PWR2 connectivity provides a practical advantage for variational learning.

\section{Applications}
\label{sec:applications}
\subsection{VQE for the Power-of-Two Transverse field Ising model}
\label{sec:vqe_pwr2}
We now assess how circuit geometry influences ground-state learning by applying the variational quantum eigensolver (VQE) to a family of long-range interacting spin models. Specifically, we consider the variable-range transverse-field Ising model characterized in Ref.~\cite{gunning_geometry-driven_2025}. This model is a transverse-field antiferromagnetic Ising Hamiltonian with algebraically weighted interactions among spins whose separations are restricted to powers of two.

The Hamiltonian takes the form
\begin{equation}
    H_{\mathrm{PWR2}} = \sum_{i=1}^{n} \sum_{d \in D}
        J_d\, \sigma^z_i \sigma^z_{i+d} + B \sum_{i=1}^{n} \sigma^x_i ,
    \label{eq:ham_isingpwr2}
\end{equation}
where $B$ denotes the transverse field and the interaction strengths
are given by $J_d = J\, d^{s}$. We assume periodic boundary conditions. The distances are drawn from the set
\begin{equation}
    D = \{ 2^l \mid l = 0, 1, \ldots, \log_2(n/2) \},
\end{equation}
which enforces a sparse, hierarchical coupling structure between spin pairs separated by a power of two. As a result, the interaction graph of the Hamiltonian coincides with the coupling graphs underlying the long-range circuits introduced earlier.

The exponent $s$ tunes the effective interaction range. For $s < -2$, the system is effectively short-range and reduces to a nearest-neighbor transverse-field Ising model. In the intermediate regime $-2 < s < 2$,
the interaction strengths are comparable across distances, and the model has been shown to exhibit behavior similar to that of an all-to-all coupled system. For $s > 2$, the Hamiltonian is strongly dominated by the most distant couplings.
We benchmark the VQE performance at the representative parameter values $s = \pm5$.

We compare several ansätze that differ only in their gate range pattern at identical circuit depth and parameter count. Our primary ansatz is a Hamiltonian variational ansatz circuit consisting of
alternating layers of single-qubit $R_X$ rotations and two-qubit $R_{ZZ}$ entanglers at a specified set of distances; this matches the gate generators of $H_{\mathrm{PWR2}}$. We report results at depths
$d \leq 3$ with gate-range patterns from nearest-neighbor only $d=1$ to the full power-of-two long-range pattern $d=8$. To verify that our findings are not specific to this ansatz, we additionally
benchmark a hardware-efficient ansatz consisting of $R_Y$ rotations followed by CNOT entanglers, and a XY ansatz with alternating $R_{XX}$, $R_{YY}$ entanglers, CNOT gates and arbitrary single-qubit rotations (results in App.~\ref{app:VQE_add}). Both an initialization in $\ket{0}^{\otimes n}$ as well as $\ket{+}^{\otimes n}$ were tested together with exchanging the order of the $R_X$ and $R_{ZZ}$ gates. The choice mattered because  $\ket{+}$ is an eigenstate of the $R_X$ gates of the first layer, which thus acts trivially. This motivates the change in the gate ordering depending on the initial state. All circuits are trained with the L-BFGS optimizer for $480$ iterations. The cost function is the variational energy,
\begin{equation}
\mathcal{C}(\bm{\theta}) =
\bra{\psi(\bm{\theta})} H_{\mathrm{PWR2}} \ket{\psi(\bm{\theta})} ,
\end{equation}
and each configuration is run with $10$ independent random seeds. We report the median relative energy error $\lvert\delta E\rvert = \lvert E_{\mathrm{VQE}} - E_{\mathrm{GS}}\rvert/\Delta$ where $\Delta$ is the spectral gap of the Hamiltonian
with interquartile range across seeds as well as the results of the individual seeds.
To provide a baseline for comparison, the reference ground-state energy $E_{\mathrm{GS}}$ and the spectral gap $\Delta$ is obtained from sparse diagonalization.

\begin{figure}
    \includegraphics[width=\linewidth]{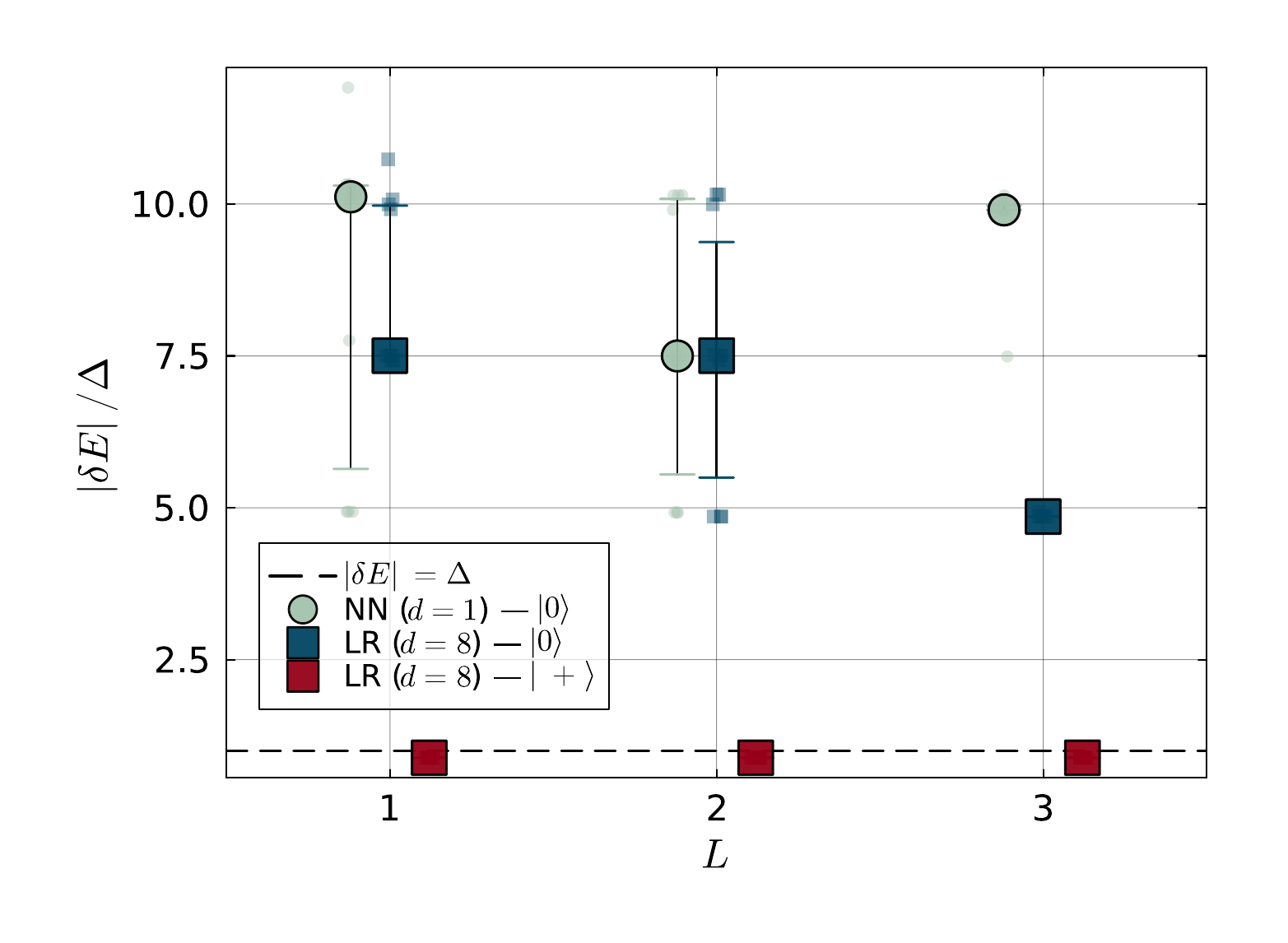}
    \caption{Median relative energy error $\lvert\delta E\rvert/\Delta = \lvert E_{\mathrm{VQE}} - E_{\mathrm{GS}}\rvert/\Delta$ as a function of the number of layers $L$, for the Hamiltonian variational $R_X$-$R_{ZZ}$ ansatz with two gate-range patterns: nearest-neighbor (NN, green circles, $d=1$) and long-range (LR, blue/red squares, $d=8$). Large filled markers are medians over $10$ random initializations with error bars indicating the interquartile range; faint markers show individual seeds. Results are  $n = 16$ qubits at $B/J = 0.05$ with $s = +5$, $\ket{0}^{\otimes n}$ and $\ket{+}^{\otimes n}$ initial states. We use the L-BFGS optimizer with $480$ iterations. For the LR ansatz, we include both initializations.
    }
    \label{fig:VQE_results_NN_vs_long}
\end{figure}

Figure~\ref{fig:VQE_results_NN_vs_long} compares the median energy error relative to the spectral gap at $s = +5$ for the nearest-neighbor and full long-range (LR) gate patterns of the HVA. We show results for depths $d\leq3$. For the LR ansatz, we present both initializations. We exclude the $|+\rangle$ results for the NN circuits, as they get trapped in a high-energy basin outside the plot range. The long-range ansatz reaches an absolute energy error of $\lvert E_{\mathrm{VQE}} - E_{\mathrm{GS}}\rvert\approx 2.7\times 10^{-3}$ at depth~$1$. This absolute energy error is two orders of magnitude below the nearest-neighbor median. In particular, the (LR, $|+\rangle$) ansatz achieves energy errors below the spectral gap $\Delta$, thereby resolving the low-energy states.

However, the energy comparison alone can be misleading. In most nearest-neighbor VQE runs at $s = +5$, the solution reproduces the local order of the ground state ($\langle \sigma_i^z \sigma_j^z\rangle \approx \pm 1$, $\langle \sigma_i^x \rangle \approx -B$), but the fidelity with the true ground state is low: $\mathcal{F}_{\mathrm{med}} \approx 0.05$.

 On inspection, the variational states found by the NN ansatz are $R_X$-tilted computational basis states, $|\psi\rangle \approx \bigotimes_i R_X(\theta_i), |b_i\rangle$. The bitstring $b$ varies by seed, and overlaps between seeds are zero. Under L-BFGS, the nearest-neighbor ansatz behaves as a classical solver over basis states. It finds a low-energy computational state with the correct magnetic order but lacks entanglement and overlap with the entangled ground state. With $\ket{0}^{\otimes n}$ initialization, even the long-range circuits stay trapped in product-state basins at all tested depths. With the $|+\rangle$ initialization, all seeds converge to an entangled state with $S(n/2) \approx 5.5$, slightly overshooting the exact value $S_{GS} \approx 5.1$.

These findings demonstrate that aligning the circuit connectivity with the Hamiltonian geometry in the long-range regime. Local observables and the variational energy alone cannot diagnose ground-state convergence. At $s = +5$, the near-degenerate low-energy spectrum places near-classical configurations only $\mathcal{O}(10^{-2})$ in energy above the ground state. As a result, the optimizer for a nearest-neighbor ansatz often settles into these configurations, missing the entangled low-energy manifold. The half-chain entanglement entropy sharply detects this collapse. In the Appendix~\ref{app:VQE_add}, we additionally show how the achieved variational fidelity scales with increasing transverse field $B$. The larger energy gap at $B=0.05$ makes the ground state resolvable by VQE, while keeping the same qualitative physics. Even at low depths, long-range ansätze can encode distant correlations within the circuit, accessing the entangled ground-state manifold. In contrast, the nearest-neighbor circuits require deeper circuits to match this expressibility, thus, providing numerical evidence for the results shown in Fig.~\ref{fig:entangling_Cap_combined}.

\subsection{Variational Learning of Global Quench Dynamics in the TFIM}
\label{sec:vqa_tfim}
As a more complex example for the PWR2 ansätze, we study a variational quantum algorithm (VQA) designed to reproduce the non-equilibrium dynamics of the one-dimensional transverse-field Ising model (TFIM) following a global quench.

The target dynamics are generated by the Hamiltonian
\begin{equation}
    H(h) = -J \sum_i \hat{\sigma}^z_i \hat{\sigma}^z_{i+1} - h \sum_i \hat{\sigma}^x_i,
    \label{eq:H_IsingNN}
\end{equation}
with $J = 1$ setting the energy scale.

The quench protocol mimics a transition from a disordered to an ordered phase, corresponding to an instantaneous change of the transverse field from $h_i = \infty$ to $h_f = 1$. This setting provides a paradigmatic benchmark for the emergence of light cones, the spreading of correlations, and the growth of entanglement in quantum spin systems.
Starting from the fully polarized initial state
$|\psi_0\rangle = |+\rangle^{\otimes n}$,
we variationally optimize a parametrized circuit $U(\boldsymbol{\theta})$ to approximate the time-evolved state
\begin{equation}
    |\psi(t)\rangle = e^{-i H(h_f)t} |\psi_0\rangle,
\end{equation}
for various evolution times $t$. The optimization minimizes the infidelity between the circuit state and the exact time-evolved state, as obtained via exact diagonalization for the system sizes considered.

\begin{figure}[t]
    \centering
    \includegraphics[width=\linewidth]{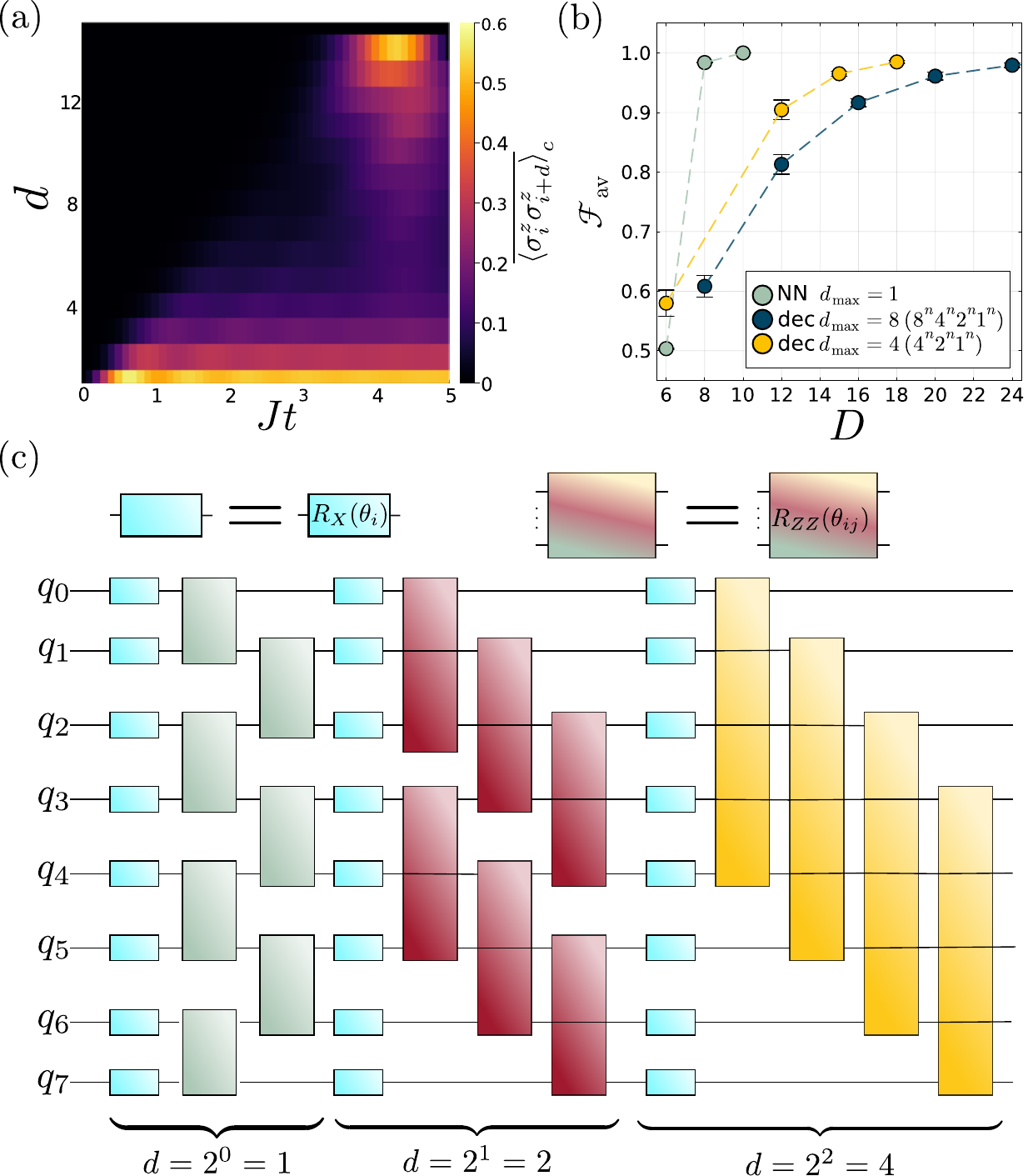}
    \caption{Learning the late-time state of a nearest-neighbor TFIM quench. (a) Connected two-point correlator $\langle \sigma^z_i\sigma^z_{i+d}\rangle_c$
  of the target state vs separation $d$ and quench time $Jt$ for the NN TFIM at $n = 16$, showing the light-cone spread of correlations.(b) Average fidelity $\mathcal{F}_{av}$ of the variational state vs total depth $D$ (in Trotter blocks of depth 2) for three ansätze: NN ($d=1$ only, green), decreasing-distance with $d_{max}=4$ (pattern $4^k2^k1^k$ yellow), and the full power-of-two pattern with $d_{max}=8,$ $8^k4^k2^k1^k$, blue). One full Trotter step corresponds to 1, 3, and 4 blocks, respectively.
(c) Schematic of the decreasing-distance ansatz on $n = 8$ illustrating the $d_{max}-4$ pattern: single-qubit $R_X$ rotations followed by $R_{ZZ}$ entanglers at $d=4,2,1$ in sequence (each color groups one distance scale).
}
\label{fig:VQA_results}
\end{figure}

We compare three ansätze that differ in their connectivity structure: a nearest-neighbor (NN) brick-wall circuit with $d=1$ couplings only; the full power-of-two pattern with $d_{\max}=8$ (the $8^n 4^n 2^n 1^n$ decreasing-distance arrangement); and a truncated PWR2 variant with
$d_{\max}=4$ that omits the longest-range coupling as shown in Fig.~\ref{fig:VQA_results}(c). We measure depth in units of Trotter blocks of two layers each, so that one block executes a complete distance-$d$ layer for either ansatz family; the full Trotter step of the NN, $d_{\max}=4$, and $d_{\max}=8$ ansätze therefore corresponds to $1$, $3$, and $4$ blocks, respectively.

Figure~\ref{fig:VQA_results}(b) summarizes the performance of the three ansätze for learning the post-quench state at $Jt = 5$. All three reach high fidelity with the exact target, but the NN ansatz does so at the smallest depth ($D \approx 10$ blocks), while the PWR2 ansätze require approximately $1.8\,\times$ ($d_{\max}=4$) and $2.4\,\times$ ($d_{\max}=8$) the depth to reach comparable fidelity. We attribute this overhead to a mismatch between the ansatz and the gate generators of the target Hamiltonian: the target dynamics is generated by a nearest-neighbor Hamiltonian and, to first order in the Trotter expansion, is therefore expressed purely in terms of nearest-neighbor gates. The PWR2 ansatz instead allocates variational capacity to gates at distances $d > 1$ that do not appear in the generator of $e^{-i H_{\mathrm{NN}} t}$, and must spend parameters to reproduce the NN-generated structure with its richer gate set. The overhead is therefore not a property of the target state's correlation length which has by $Jt=5$ spread across the entire chain (see Fig.~\ref{fig:VQA_results}(a)) but of the mismatch
between the ansatz's connectivity and the connectivity of the target Hamiltonian.

We also observe a performance dependence on layer ordering which was already noted in~\cite{lyu_variational_2023}: empirically, the best ordering groups the layers by distance and places the most dominant interaction length scale at the end of the circuit. We confirm this finding, with the decreasing-distance ordering ($8^n 4^n 2^n 1^n$) consistently outperforming alternative arrangements of the same gates (data not shown).

The expected advantage of the PWR2 graph for variational quench compression is asymptotic and lies in how the required depth scales with system size. A single Trotter block on the NN graph implements one full Trotter step, and the scrambling time of the NN chain is set by the Lieb--Robinson velocity, suggesting $D_{\mathrm{NN}} = \mathcal{O}(n)$ for
spanning the chain. On the PWR2 graph, a single block implements only one of the $\log_2(n)$ coupling distances, and the scrambling time of the graph is itself logarithmic due to its expander-like structure, suggesting $D_{\mathrm{PWR2}} = \mathcal{O}(\log^2(n))$. These estimates are heuristic and are based on graph diameter and information-propagation arguments rather than a rigorous analysis of variational optimization complexity.

The scrambling time can be understood by considering the PWR2 graph on $n = 2^k$ qubits which  has edges at $\log_2 n$ distinct distances, $d \in \{2^0, 2^1, \ldots, 2^{k-1}\}$. For any displacement $\Delta = |i - j| \in \{1, \ldots, d-1\}$ between two sites, the binary expansion
\begin{equation}
    \Delta = \sum_{j=0}^{n-1} b_j\, 2^j,\qquad b_j \in \{0,1\},
\end{equation}
identifies an explicit path of length at most $\log_2 n$ from $i$ to $j$ that hops along one edge per non-zero bit of $\Delta$. The diameter of the PWR2 graph is therefore $\log_2 n$, in contrast to the linear diameter of the NN chain. Combined with the fact that a single Trotter step on the PWR2 graph requires $\log_2 n$ blocks (one per distance
layer), this $\log n$ fan-out yields the $\mathcal{O}(\log^2 n)$ scaling of $D_{\mathrm{PWR2}}$ reported above: one factor counts the number of Trotter steps required for the causal cone to span the system, the other counts the depth of a single Trotter step. Extrapolating the empirical prefactors obtained at $n=16$, suggests a  cross over near $n \approx 100$--$130$. If the observed trends persist at larger system sizes, the PWR2 ansatz may require fewer Trotter blocks than its NN counterpart at the comparable target fidelity. The result at $n=16$ should therefore be read as a controlled demonstration that the PWR2 ansatz can learn the target state with high fidelity, supporting the possibility that the asymptotic advantage of the PWR2 connectivity is not undermined by an expressibility gap at the small system sizes we can directly simulate.

\section{Combining variational learning with reconfigurability}
\label{sec:monnamap}
In addition to incorporating long-range entangling gates directly into the variational ansatz, long-range correlations can also be addressed by transforming the system’s underlying representation. We introduce a Monna-map–based framework that maps extended interactions to local ones, enabling variational optimization with short-range circuits while preserving access to long-range physics via postprocessing.

The Monna map was introduced in \cite{monna_sur_1952} and later discussed in the context of simulating the dynamics of tree graphs \cite{bentsen_treelike_2019,hashizume_deterministic_2021,gunning_geometry-driven_2025} by transforming between a linear and a treelike representation of a coupling graph. For an illustration, consider the coupling graph underlying the PWR2 model at $s=5.0$, i.e., the limit in which furthest neighbor interactions are the strongest.  Applying the Monna map to the PWR2 graph and the transverse-field Ising model in the limit $s=5.0$, the transformation maps the most strongly interacting bonds from furthest neighbors to nearest neighbors.  More generally, the Monna map induces a hierarchical (tree-like) reordering of sites, rendering dominant long-range couplings local. We can use this reordering to obtain a representation of the interaction Hamiltonian in which dominant couplings act locally. This is, for example, useful when applying MPS methods, as it reduces the number of bonds with high bond dimension, which is the main factor limiting scalability, but it can also be exploited in variational learning, as  in the following. We can write the Monna map as a unitary $U_\mathcal{M}$, which can be implemented at the circuit level as a series of swap operations between the qubits. We can apply the map to quantum states as well as to operators. Applying this to the PWR2 Ising Hamiltonian, i.e., $H_\mathcal{M}=U_\mathcal{M}HU_\mathcal{M}^\dagger$, leaves its spectrum invariant, since the two Hamiltonians are unitarily equivalent. This means that, in particular, the ground state energies are the same. We can exploit this fact for variational learning. Consider that we learn with respect to a cost function linear in the density matrix. We can use the cyclicity of the trace to note that $C(\boldsymbol{\theta})=\mathrm{Tr}(\rho_{\boldsymbol{\theta}}\hat{O}) = \mathrm{Tr}(U_\mathcal{M} \rho_{\boldsymbol{\theta}} U_\mathcal{M}^\dagger  U_\mathcal{M} \hat{O} U_\mathcal{M}^\dagger)$. For the choice of $\hat{O}=\hat{H}$, we recover the cost function, minimizing the energy in VQE. The value of the cost function will then be the same for both Hamiltonians and the corresponding states in the original and mapped frames. Combining unitarity of the Monna map with the locality of  the mapped Hamiltonian $H_\mathcal{M}$ and restricting  to simple product initial states, such that  $U_\mathcal{M}|\sigma_1\cdots\sigma_N\rangle=|\sigma_{\mathcal{M}(1)}\cdots \sigma_{\mathcal{M}(N)}\rangle$  we can efficiently apply the Monna map to the  state classically, and arrive at the following procedure for preparing the ground state of a long-range correlated Hamiltonian (see also Fig.~\ref{fig:overview}(b)). We note that the idea of using permutation in variational ansaetze was independently introduced and developed in \cite{tkachenko_correlation-informed_2021} in the context of VQE for small molecular systems. There, the permutations were introduced to localize the mutual information in the approximated ground state. The optimal permutation is found via a preprocessing step and was shown to reduce the necessary circuit depth to reach chemical accuracy efficiently compared to using the linear encoding of orbitals into qubits. In contrast, our motivation acts on the level of the hamiltonian interaction graph. In the appendix \ref{app:monna}, we show however that it in the case of our example the Monna map also localizes the mutual information in exact ground state.

We transform the Hamiltonian under the Monna map, $H_\mathcal{M}=U_\mathcal{M}HU_\mathcal{M}^\dagger$. We learn the ground state of the transformed Hamiltonian with a short-range/NN variational circuit and with respect to   the variational energy $C(\boldsymbol{\theta})=\langle\psi(\boldsymbol{\theta})|H_\mathcal{M}|\psi(\boldsymbol{\theta)\rangle}$.  After the optimization, we obtain the best approximation to the ground state of the transformed Hamiltonian by applying the circuit with the optimized parameters to the initial state.  The resulting state now has the same energy with respect to both Hamiltonians, i.e. $\langle \psi | H_\mathcal{M} | \psi \rangle = \langle \psi_{\text{phys}} | H | \psi_\text{phys} \rangle$. To recover the true ground state, we must apply the inverse Monna map  $U^\dagger_\mathcal{M}$ to the optimized state. Note that standard VQE bounds based on spectral gaps for the accuracy of correlations or observables are derived from the operator spectrum and hence carry over in this scheme. i.e., if we can learn the ground state of the mapped Hamiltonian, we can prepare the ground state of the original Hamiltonian. In a MPS representation, SWAP operations between distant sites are not local. Each SWAP must be decomposed into a sequence of nearest-neighbor exchanges, and when applied across an entanglement cut, this generically increases the Schmidt rank, leading to a rapid growth of the required bond dimension and rendering the inverse Monna map classically inefficient and infeasible for larger system sizes. However, we can switch to a quantum device on which the optimized state can be prepared using the optimized circuit. Subsequently, one can apply SWAP operations or physically reconfigure the atoms to recover the original ground state. Here, the application of the inverse of the Monna map corresponds to an atom-shuttling operation, which can be implemented in neutral-atom devices with reconfigurable tweezer arrays, where physical qubit rearrangement is a native operation. Once the many-body ground state is prepared, i.e., first apply the circuit and subsequently the shuttling, it is possible to perform subsequent quantum operations of interest, such as computing quantities that are non-linear in the density matrix, e.g., entanglement entropies.

Alternatively, the transformation can be applied to the optimized circuit, yielding a highly non-local unitary. One can then apply this transformed, optimized circuit to the initial state. This will equivalently yield the true ground state. This, however, will have higher complexity scaling as  $\mathcal{O}(d^2)$ (where $d=2^n$ is the Hilbert space dimension, rendering this approach classically infeasible for large $n$) and relies on the ability to perform high-fidelity long-range gates. Whether this is preferable to the preceding method depends on the performance of the specific quantum device.

The Monna map permutes the $2^n$ computational basis states by bit reversal. This permutation decomposes into $\frac{2^n-2^{\lceil n/2\rceil}}{2}$ disjoint transpositions. In an abstract circuit model, it can be implemented with $\mathcal{O}(n)$ SWAP operations. On linear-array hardware with only nearest-neighbor connectivity, each transposition may require $\mathcal{O}(n)$ adjacent SWAP gates. This results in an overall gate count of $\mathcal{O}(n^2)$ and a circuit depth of $\mathcal{O}(n)$ if swaps are parallelized in a brickwork pattern. In contrast, platforms with programmable geometry, such as neutral-atom tweezer arrays, can implement the permutation by physically rearranging atom positions or reassigning logical indices. This eliminates the need for additional coherent SWAP gates, reducing the coherent circuit overhead of the permutation to constant depth. Thus, the Monna mapping is well-suited to architectures that use classical control rather than unitary operations for qubit reordering. We note that while the coherent circuit overhead is reduced to constant depth, the shuttling operations themselves incur a classical time cost and are subject to device-specific fidelity constraints.

\begin{figure}[!t]
    \centering
    \includegraphics[width=\linewidth]{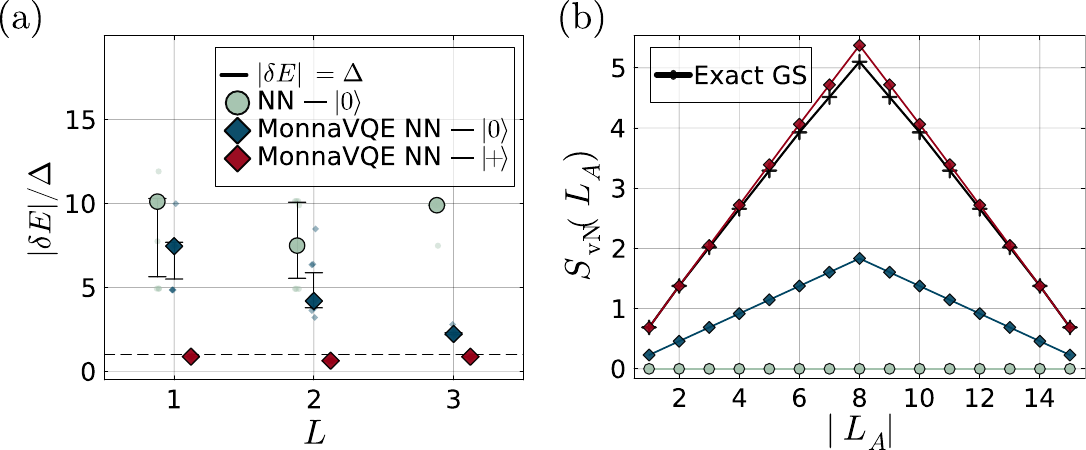}
    \caption{Monna-mapped VQE results for $n=16$ qubits at $s = 5$. Short-range NN ansätze fail when trained directly on the physical Hamiltonian $H$, but succeed when trained on the Monna-mapped Hamiltonian $H_\mathcal{M}$. (a)~Median relative energy error $\lvert\delta E\rvert/\Delta = \lvert E_\mathrm{VQE} - E_\mathrm{GS}\rvert/\Delta$ as a function of the number of layers $L$ for direct NN VQE(green circles) with $\ket{0}^{\otimes n}$ initialization and MonnaVQE (diamonds) NN for VQE with $\ket{0}^{\otimes n}$ (blue) and $\ket{+}^{\otimes n}$ (red) initialization. Large markers are medians over $10$ seeds with interquartile error bars; faint dots show individual seeds.(b)~Averaged  von-Neumann entanglement entropy $\langle S_\mathrm{vN}\rangle$ of the optimized variational state for the best seed at $L=2$. The black line marks the exact ground-state entropy $S_\mathrm{GS} = 5.10$, characteristic of a volume law state. The direct VQE pipeline collapses to a product state ($\langle S_\mathrm{vN}\rangle = 0$  while MonnaVQE builds the entanglement structure required to access the entangled low-energy manifold.
}
    \label{fig:monna_results}
\end{figure}

We validate the Monna-mapped variational pipeline by demonstrating that ground states with strong long-range correlations can be learned using strictly short-range variational circuits, without introducing any approximation at the level of the Hamiltonian or the measured observables. As a benchmark, we consider variational ground-state learning for the long-range transverse-field Ising model in the strongly long-range regime $s=5$ at magnetic field $B=0.05$ (spectral gap $\Delta=3.0\times10^{-3}$), using the same cost function and optimization protocol as in Section~\ref{sec:vqe_pwr2}. All simulations are performed for $n=16$ spins. Both the direct and Monna pipelines use the identical NN HVA ansatz (16 $R_x$ + 16 NN $R_{ZZ}$ gates per layer) with matched circuit depth and identical numbers of variational parameters, and L-BFGS optimization for $480$ iterations with $10$ independent random seeds at each depth from $1$ to $3$. The only difference between the two pipelines is whether the NN gates act on the physical qubit register (direct VQE) or on the qubit register obtained by composing the bit-reversal Monna permutation $\sigma$ with the physical assignment (MonnaVQE). We test two initializations: $|0\rangle^{\otimes n}$ with $R_X$-first gate ordering, and $|+\rangle^{\otimes n}$ with $R_{ZZ}$-first ordering. 

Figure~\ref{fig:monna_results}(a) shows the median relative energy error $|\delta E|/\Delta$ as a function of circuit depth for the direct and Monna pipelines under both initializations. The direct NN pipeline is trapped far above the ground state regardless of initialization: under $|0\rangle^{\otimes n}$ it converges to a product-state basin at $|\delta E|/\Delta\approx7$–$10$ across all depths, while under $|+\rangle^{\otimes n}$ it fails catastrophically ($|\delta E|/\Delta>2000$). The Monna pipeline, in contrast, succeeds under both initializations. With $|0\rangle^{\otimes n}$, the median error decreases monotonically from $\sim7.5,\Delta$ at depth $1$ to $\sim2.3,\Delta$ at depth $3$. With $|+\rangle^{\otimes n}$, all seeds converge to the same state with zero variance, reaching $|\delta E|/\Delta=0.64$ at depth $2$ below the spectral gap. 

The contrast sharpens upon inspection of the variational states themselves. Figure~\ref{fig:monna_results}(b) shows the cyclically averaged von Neumann entanglement entropy $\langle S_{vN}(L_A)\rangle$ as a function of subsystem size for the highest-fidelity seed at depth $2$, with the exact ground-state profile ($S_{GS}(n/2)=5.10$) shown as a reference. The direct NN pipeline under $|0\rangle^{\otimes n}$ produces a product state with vanishing entanglement at all bipartition sizes, confirming that it is trapped in a classical energy minimum. The Monna pipeline under $|+\rangle^{\otimes n}$ closely reproduces the exact Page curve, with $S_{vN}(n/2)=5.37$ and fidelity $\mathcal{F}=0.85$, overshooting the ground-state entropy by only $0.27$ bits. Under $|0\rangle^{\otimes n}$, the Monna pipeline builds partial entanglement ($S_{vN}(n/2)=1.83$, $\mathcal{F}=0.06$), demonstrating that even with a suboptimal initialization, the permuted register accesses the entangled low-energy manifold that the direct pipeline cannot reach. 

The Monna mapping, therefore, eliminates the product-state attractor that traps the direct pipeline, not by increasing the circuit’s expressibility, but by aligning the locality of the entangling gates with the dominant couplings of the Hamiltonian. The same nearest-neighbor gate set, applied to a permuted qubit register, generates the long-range entanglement that the physical-register assignment cannot. The $|+\rangle^{\otimes n}$ initialization yields the stronger result because it provides a translationally symmetric starting point with finite overlap onto the entangled ground-state manifold; however, the improvement of the Monna pipeline under $|0\rangle^{\otimes n}$ shows that the advantage is not contingent on a favorable initialization.

 The advantage of the Monna-mapped approach depends on the interaction structure of the target Hamiltonian and is most pronounced for sparse, hierarchical long-range coupling graphs. At the uniform point $s=0$, where every pair of sites couples with the same strength, no qubit permutation can simultaneously localize all couplings, and the mapping yields no significant benefit. In the strongly short-range regime $s=-5$, the ground state is near-classical, and the NN ansatz already reaches the ground-state energy without requiring long-range entanglement (Section\ref{sec:vqe_pwr2}). The Monna mapping, therefore, acts as a problem-aware classical preprocessing step: in regimes where the ground state is highly entangled, and the entanglement is organized hierarchically by the coupling graph, it turns a structurally limited optimization into a feasible one. The preprocessing cost is purely classical (a permutation of the qubit register) and, on platforms with reconfigurable connectivity, such as neutral-atom tweezer arrays, can be implemented with constant-depth quantum overhead.

\section{Discussion and Outlook}
\label{sec:discussion}

In this work, we evaluated the potential of circuits with PWR2 connectivity, comparing them to circuits with limited NN connectivity at comparable circuit depth and parameter count. We demonstrated this in two cases: a model whose structure mirrors the circuit connectivity exactly, and a geometrically local model in which long-range correlations emerge only during late-time dynamics. As always, there is a tradeoff between the expressibility and trainability of the ansatz. On the expressibility side, there is an advantage to long-range coupling graphs beyond small system sizes, as they can capture long-range correlations with fewer layers than are required for local coupling geometries. This is naturally dependent on the geometry of correlations in the given model. 

In particular, we note that structured long-range connectivity offers a practical middle ground between local and fully connected architectures. Using a sparse graph improves the ability to represent long-range correlations at fixed circuit depth because of the addition of long-range connectivity relative to locally connected architectures, while requiring fewer connections (and therefore layers) per qubit than fully connected architectures.

The central open question is whether the potential performance advantages of PWR2 connectivity tested here remain for classically unverifiable system sizes, a question our numerical results, restricted to $n = 16$ qubits, cannot yet resolve. Classical computation becomes challenging beyond this scale, in part because matrix product state methods are inefficient for long-range-coupled models. Separately, our results show that gate-order dependence and ansatz-dependent qubit permutations are concrete mechanisms through which circuit geometry shapes the optimization landscape, and understanding these interactions is essential for deploying these methods at scale.

These questions could be directly addressed on near-term hardware. The methods here are especially relevant to current and near-term neutral-atom quantum processors using optical tweezer arrays \cite{saffman_quantum_2010,saffman_quantum_2016,bluvstein_logical_2024,finkelstein_universal_2024,muniz_high-fidelity_2025,chiu_continuous_2025,bluvstein_fault-tolerant_2026} and trapped ions using collective modes \cite{monroe_programmable_2021,foss-feig_progress_2025,ransford_98-qubit_2026}. In these platforms, long coherence times and reconfigurable connectivity enable both structured long-range entangling gates and qubit reordering via atom rearrangement. As a result, circuit geometry becomes a design parameter rather than a fixed hardware constraint, providing the opportunity to compare the differently structured ans\"atze we have considered here at larger size scales. 

A further consequence of this geometric flexibility in neutral atom arrays is efficient reconstruction of a target geometry after variational optimization with permuted qubits. Whether implemented via atom rearrangement or logical reindexing of qubit labels, this step does not significantly increase circuit depth, limiting the addition of coherent gate errors. The effectiveness of this approach depends on both the problem's sparsity and whether a suitable permutation is known.

Combining structured long-range ansätze with Monna-mapped locality discussed here opens several concrete research directions. For example, the locality structure of the mapped representation could be exploited to extend variational learning to larger system sizes. Gate ordering and connectivity could also be optimized adaptively. Geometry-aware regularization schemes could further improve optimization. Together, these directions suggest that connectivity and gate ordering deserve the same systematic attention as circuit depth in the design of variational quantum algorithms.

\acknowledgements
 Numerical computations used the Yao.jl package \cite{luo_yaojl_2020}. We thank Matthias C. Caro and Alex Gunning for helpful discussions. This work was supported by the EPSRC through the QQQS programme grant (EP/Y01510X/1) and by the QCi3 hub (EP/Z53318X/1).
 
\bibliography{bibliography}

\appendix
\section{Proof of Commutator depth bound }
\label{app:A}

Here we answer the finite depth question, asking how many commutator layers are required before the PWR2-generated filtration already contains the nearest-neighbor TFIM target algebra. In the following, we track the quadratic Majorana sector generated by the corresponding edge supports. Thus the result is a hitting-depth bound for the algebra $M_{2n}\cong\mathfrak{so}(2n)$ starting from different generator sets.

Let
\begin{equation}
M_{2n}:=\operatorname{span}\{M_{ab}:1\leq a<b\leq 2n\},
\qquad
M_{ab}:=\frac{1}{2}\gamma_a\gamma_b .
\end{equation}
For an edge set $E$, define
\begin{equation}
L_0(E):=\operatorname{span}\{M_{ab}:(a,b)\in E,\ a<b\},
\end{equation}
and the nested commutator
\begin{equation}
L_{q+1}(E):=\operatorname{span}\left(L_q(E)\cup [L_q(E),L_0(E)]\right).
\end{equation}
This equation tracks the quadratic Majorana sector associated with the edge supports $E$.

For the open nearest-neighbor edge set
\begin{equation}
E^{\rm nn}:=\{(a,a+1):1\leq a<2n\},
\end{equation}
the least depth $q^{\rm nn}_{\mathfrak{so}(2n)}$ such that
\begin{equation}
L_{q^{\rm nn}_{\mathfrak{so}(2n)}}(E^{\rm nn})=M_{2n}
\end{equation}
is
\begin{equation}
q^{\rm nn}_{\mathfrak{so}(2n)}=2n-2.
\end{equation}

For the open power-of-two edge set
\begin{equation}
E^{\rm pwr2}:=\{(a,a+2^r):1\leq a<a+2^r\leq 2n,\ r\geq 0\},
\end{equation}
the least depth $q^{\rm pwr2}_{\mathfrak{so}(2n)}$ such that
\begin{equation}
L_{q^{\rm pwr2}_{\mathfrak{so}(2n)}}(E^{\rm pwr2})=M_{2n}
\end{equation}
satisfies
\begin{equation}
q^{\rm pwr2}_{\mathfrak{so}(2n)}
\leq
\left\lfloor \log_2(2n-1)\right\rfloor .
\end{equation}

To show this we use the standard quadratic Majorana commutator identity
\begin{equation}
[M_{ab},M_{cd}]
=
\delta_{bc}M_{ad}
-\delta_{ac}M_{bd}
-\delta_{bd}M_{ac}
+\delta_{ad}M_{bc}.
\end{equation}
In particular, if $a<b<c$, then
\begin{equation}
[M_{ab},M_{bc}]=M_{ac}.
\end{equation}

Let $G=(V,E)$ be the graph associated with the edge set $E$, and let $\operatorname{dist}_G(a,b)$ denote graph distance. We first note that the nested commutators satisfy
\begin{equation}
L_q(E)
=
\operatorname{span}\{M_{ab}:1\leq a<b\leq 2n,\ \operatorname{dist}_G(a,b)\leq q+1\}.
\end{equation}
Indeed, $L_0(E)$ contains precisely the bilinears supported on graph edges. If $a=x_0,x_1,\ldots,x_m=b$ is a shortest path with $m\leq q+2$, then by induction $M_{x_0x_{m-1}}\in L_q(E)$ and $M_{x_{m-1}x_m}\in L_0(E)$. Their commutator gives $\pm M_{ab}$, so $M_{ab}\in L_{q+1}(E)$. 

Conversely, a commutator with an element of $L_0(E)$ can extend a path by at most one graph edge. Therefore the first depth at which all bilinears are generated is
\begin{equation}
q_*(E)=\operatorname{diam}(G)-1.
\end{equation}

For $E^{\rm nn}$, the graph is the path graph with $2n$ vertices. Its diameter is
\begin{equation}
\operatorname{diam}(G^{\rm nn})=2n-1.
\end{equation}
Hence
\begin{equation}
q^{\rm nn}_{\mathfrak{so}(2n)}
=
2n-2.
\end{equation}

For $E^{\rm pwr2}$, fix $a<b$ and set $d=b-a$. 
\begin{equation}
d=\sum_{j=1}^k 2^{r_j},
\qquad
r_1>r_2>\cdots>r_k\geq 0,
\qquad
k=\mathrm{wt}_2(d).
\end{equation}
where $\mathrm{wt}_2$ denotes the binary Hamming weight.
Define $s_0:=0$ and $s_t:=\sum_{j=1}^t2^{r_j}$. Then
\begin{equation}
a+s_0,\ a+s_1,\ldots,\ a+s_k=b
\end{equation}
is a path in $G^{\rm pwr2}$ of length $k$, since each step has power-of-two length. Therefore
\begin{equation}
M_{ab}\in L_{\mathrm{wt}_2(b-a)-1}(E^{\rm pwr2}).
\end{equation}
Since $1\leq b-a\leq 2n-1$,
\begin{equation}
\mathrm{wt}_2(b-a)
\leq
\left\lfloor \log_2(2n-1)\right\rfloor+1.
\end{equation}
Thus all bilinears $M_{ab}$ are generated once
\begin{equation}
q\geq \left\lfloor \log_2(2n-1)\right\rfloor,
\end{equation}
which proves
\begin{equation}
q^{\rm pwr2}_{\mathfrak{so}(2n)}
\leq
\left\lfloor \log_2(2n-1)\right\rfloor .
\end{equation}

\section{Second-moment operators and Weingarten calculus}
\label{app:weingarten}
In this Appendix, we provide the explicit expressions for the second-moment operators used in the main text and briefly review the Weingarten calculus underlying their analytic evaluation for the Haar measure. For a full general introduction see \cite{mele_introduction_2024} and the supplemental material of \cite{ragone_lie_2024} for this specific application.

\subsection{Second-moment operators}
Let $\mathcal{H} \simeq \mathbb{C}^d$ be a finite-dimensional Hilbert space. For an ensemble of unitaries $\mathcal{E}$ on $\mathcal{H}$, the second-moment operator is defined as
\begin{equation}
\mathcal{M}^{(2)}_{\mathcal{E}}
=
\mathbb{E}_{U \sim \mathcal{E}}
\left[
U^{\otimes 2} \otimes (U^\dagger)^{\otimes 2}
\right].
\end{equation}
Viewed as a superoperator via vectorization, it acts on operators on $\mathcal{H}^{\otimes 2}$. This operator fully characterizes the second moments of the ensemble and plays a central role in the theory of approximate unitary $2$-designs.
For the Haar ensemble on $\mathrm{U}(d)$, the corresponding operator$\mathcal{M}^{(2)}_{\mathrm{Haar}}$ can be evaluated analytically using representation theory and Weingarten calculus, as reviewed below.
We define the deviation operator for the circuit ensemble as
\begin{equation}
A^{(2)}_{\mathcal{E}}
=
\mathcal{M}^{(2)}_{\mathcal{E}} - \mathcal{M}^{(2)}_{\mathrm{Haar}}.
\end{equation}
Its spectral properties control the convergence of repeated applications of $\mathcal{E}$ toward Haar randomness. Subtracting $\mathcal{M}^{(2)}_{\mathrm{Haar}}$ removes the trivial fixed-point subspace, so that every singular value of $A^{(2)}_{\mathcal{E}}$ is non-trivial. In particular, the largest singular value
\begin{equation}
\lambda_{\max}(\mathcal{E}) := \big\| A^{(2)}_{\mathcal{E}} \big\|_\infty
\end{equation}
determines the depth required to approach an approximate unitary $2$-design, as shown in~\cite{ragone_lie_2024}. Later subsections will specialize this formalism to ensembles of brickwork circuits.

\subsection{Weingarten calculus}

The Weingarten calculus provides a systematic method to compute integrals of products of matrix elements over the unitary group with respect to the Haar measure. For $k$ copies, one has the general identity~\cite{mele_introduction_2024}
\begin{align}
&\int_{\mathrm{U}(d)} \!\!\mathrm{d}U \,
U_{i_1 j_1} \cdots U_{i_k j_k}\,
\overline{U}_{i'_1 j'_1} \cdots \overline{U}_{i'_k j'_k} \\
&=\sum_{\sigma,\tau \in S_k}
\delta_{i_1 i'_{\sigma(1)}} \cdots \delta_{i_k i'_{\sigma(k)}}\,
\delta_{j_1 j'_{\tau(1)}} \cdots \delta_{j_k j'_{\tau(k)}}\,
\mathrm{Wg}_d(\sigma^{-1}\tau), \nonumber
\end{align}
where $S_k$ is the symmetric group and $\mathrm{Wg}_d(\pi)$ denotes the Weingarten function associated with the permutation $\pi \in S_k$.

For $k=2$, only the identity permutation $\mathbbm{1}$ and the transposition
$(12)$ appear, and the corresponding Weingarten functions are given by
\begin{equation}
\mathrm{Wg}_d(\mathbbm{1}) = \frac{1}{d^2 - 1},
\qquad
\mathrm{Wg}_d((12)) = -\frac{1}{d(d^2 - 1)}.
\end{equation}

Using the $k=2$ Weingarten functions, the second-moment operator of the Haar ensemble can be written compactly in terms of permutation operators acting on
$\mathcal{H}^{\otimes 2}$. Denoting by $\mathbb{S}$ the swap operator on $\mathcal{H}^{\otimes 2}$,
\begin{equation}
\mathbb{S} \, |\psi \otimes \phi\rangle = |\phi \otimes \psi\rangle,
\end{equation}
the Haar second-moment operator takes the form
\begin{equation}
\mathcal{M}^{(2)}_{\mathrm{Haar}}
=
\frac{1}{d^2 - 1}\left(
\mathbb{I} \otimes \mathbb{I}
+ \mathbb{S} \otimes \mathbb{S}
- \frac{1}{d}\left(
\mathbb{I} \otimes \mathbb{S} + \mathbb{S} \otimes \mathbb{I}
\right)
\right).
\end{equation}
Equivalently, $\mathcal{M}^{(2)}_{\mathrm{Haar}}$ can be expressed as a projector onto the symmetric and antisymmetric subspaces of $\mathcal{H}^{\otimes 2}$ with appropriate weights.

\subsection{Second-moment operator for two-qubit gates}
Specializing the Haar second-moment operator to two-qubit gates drawn from $\mathrm{SU}(4)$, the $k=2$ moment operator $\mathcal{M}^{(2)}_{\mathrm{SU}(4)}$
acts on operators on $(\mathbb{C}^2)^{\otimes 4}$ and can be expressed, in the operator basis $\{\mathbb{I}\otimes\mathbb{I},\,\mathbb{I}\otimes\mathbb{S},\,
\mathbb{S}\otimes\mathbb{I},\,\mathbb{S}\otimes\mathbb{S}\}$, as the $4\times4$ matrix
\begin{equation}
\mathcal{M}^{(2)}_{\mathrm{SU}(4)}=
\begin{pmatrix}
1   & 0 & 0 & 0   \\
2/5 & 0 & 0 & 2/5 \\
2/5 & 0 & 0 & 2/5 \\
0   & 0 & 0 & 1
\end{pmatrix},
\end{equation}
obtained via the Weingarten calculus (see~\cite{ragone_lie_2024,mele_introduction_2024}). This operator governs the mixing of second moments induced by a single two-qubit gate.

\subsection{Embedding into an $n$-qubit system}
To describe circuit layers acting on an $n$-qubit register, we embed the two-qubit moment operator into the full operator space. For a gate acting on a pair of qubits $(i,j)$ with $i<j$, we define
\begin{equation}
\mathcal{M}^{(2)}_{\mathrm{SU}(4),(i,j)}
:=
\mathrm{Emb}_{(i,j)}\!\big(\mathcal{M}^{(2)}_{\mathrm{SU}(4)}\big),
\end{equation}
where $\mathrm{Emb}_{(i,j)}$ places the two tensor legs of $\mathcal{M}^{(2)}_{\mathrm{SU}(4)}$ on sites $i$ and $j$ and acts as the identity superoperator $\mathbbm{1}_k$ on every remaining site $k \notin \{i,j\}$. This embedding is defined for an arbitrary pair $(i,j)$, adjacent or not: introducing the permutation $\Pi_{(i,j)}$ of tensor factors that relabels sites $(i,j)$ to positions $(1,2)$ while preserving the order of the remaining sites, it can be written explicitly as
\begin{equation}
\mathcal{M}^{(2)}_{\mathrm{SU}(4),(i,j)}
=
\Pi_{(i,j)}^{-1}
\left(
\mathcal{M}^{(2)}_{\mathrm{SU}(4)}
\otimes
\mathbbm{1}^{\otimes (n-2)}
\right)
\Pi_{(i,j)} .
\end{equation}
For nearest-neighbor pairs $(i,i+1)$ the relabelling $\Pi_{(i,j)}$ is trivial and the expression reduces to the contiguous tensor product $\mathbbm{1}_1 \otimes \cdots \otimes \mathbbm{1}_{i-1} \otimes \mathcal{M}^{(2)}_{\mathrm{SU}(4)} \otimes \mathbbm{1}_{i+2} \otimes \cdots \otimes \mathbbm{1}_n$, whereas for longer-range pairs the conjugation by $\Pi_{(i,j)}$ accounts for the non-adjacent support.
A brickwork sublayer at distance $\ell$ consists of a set $P_\ell = \{(i,i+\ell)\}$ of disjoint qubit pairs. Because all gates in a sublayer act on disjoint supports, the corresponding second-moment operator factorizes as
\begin{equation}
\mathcal{M}^{(2)}_{P_\ell}
=
\bigotimes_{(i,i+\ell)\in P_\ell}
\mathcal{M}^{(2)}_{\mathrm{SU}(4),(i,i+\ell)}
\;\otimes\;
\bigotimes_{k \notin \bigcup_{(i,i+\ell)\in P_\ell}\{i,i+\ell\}}
\mathbbm{1}_k .
\end{equation}

A single physical circuit layer consists of a sequence of such sublayers applied in a fixed temporal order, corresponding to different interaction distances. Denoting the ordered set of distances by $\{\ell_1,\ldots,\ell_K\}$, the second-moment operator of the single-layer ensemble $\mathcal{E}_1$ is given by the composition
\begin{equation}
\mathcal{M}^{(2)}_{\mathcal{E}_1}
=
\mathcal{M}^{(2)}_{P_{\ell_K}}
\circ
\mathcal{M}^{(2)}_{P_{\ell_{K-1}}}
\circ
\cdots
\circ
\mathcal{M}^{(2)}_{P_{\ell_1}} .
\end{equation}
For nearest-neighbor circuits, $K=1$, while for power-of-two circuits
$K=\log_2 n$, reflecting the number of distinct interaction distances.

\subsection{Numerical evaluation}
Using the above construction, we compute $\lambda_{\max}$ numerically for nearest-neighbor and power-of-two circuit geometries and system sizes up to
$n=16$ qubits. We explicitly construct the second-moment operators for a single circuit layer and the corresponding Haar ensemble as sparse matrices.
The largest non-trivial singular value is then obtained via a diagonalization of the deviation operator. Since the PWR2 ensemble is not geometrically local,
we cannot use MPS methods, as there is no efficient MPS representation of the corresponding eigenstate.

\section{Ansätze}
\label{app:ansatze}
\begin{figure}[p]
\centering
\subfloat[Circuit 15]{%
  \includegraphics[width=0.45\columnwidth]{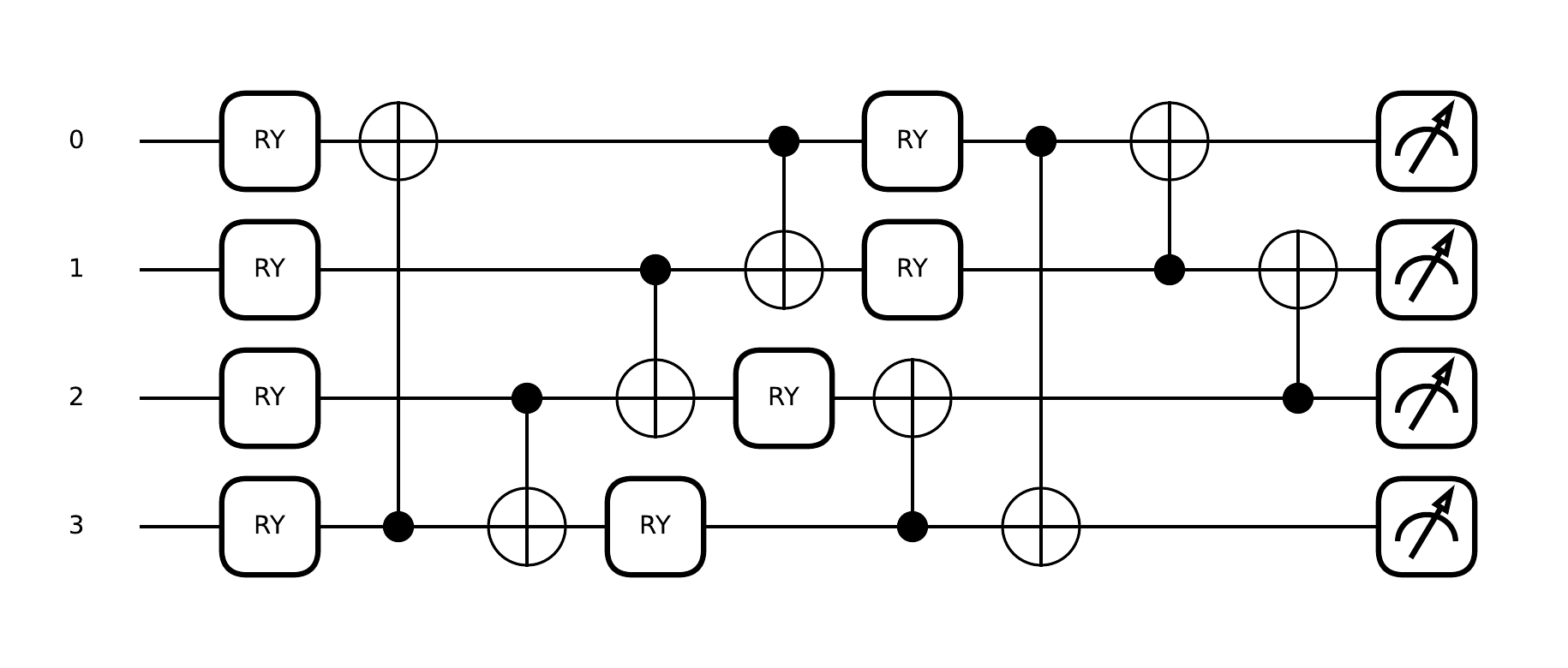}}
\hfill
\subfloat[Circuit 16]{%
  \includegraphics[width=0.45\columnwidth]{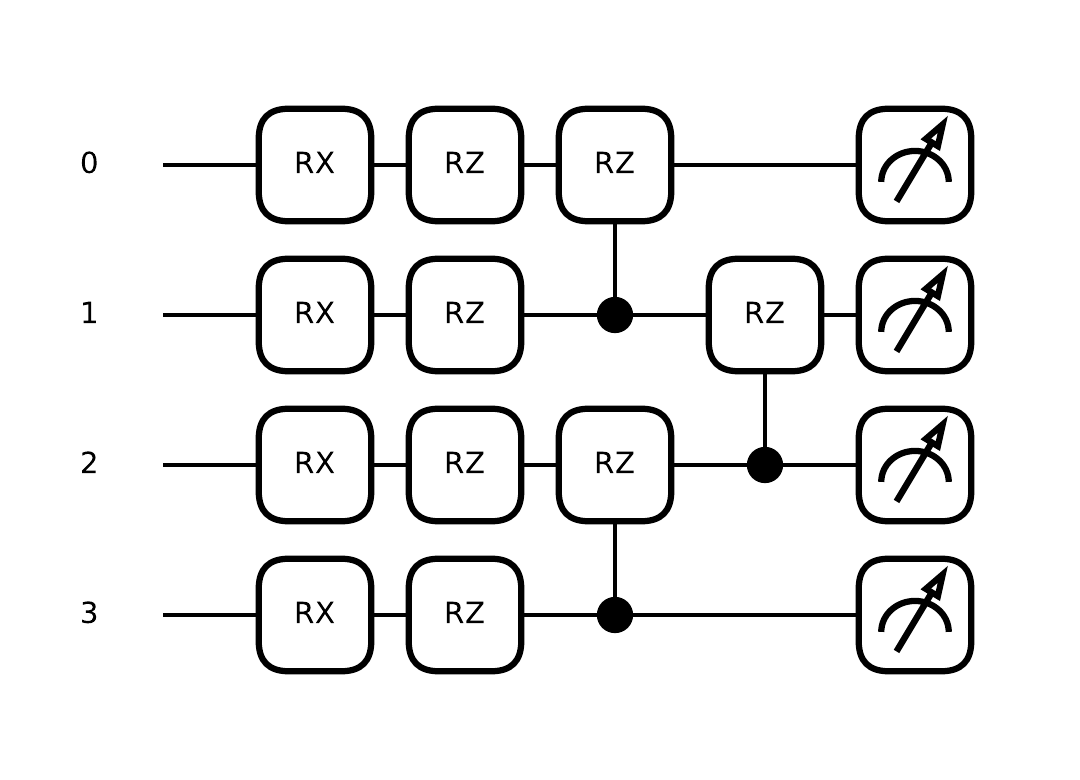}}

\vspace{0.3em}

\subfloat[Circuit 17]{%
  \includegraphics[width=0.45\columnwidth]{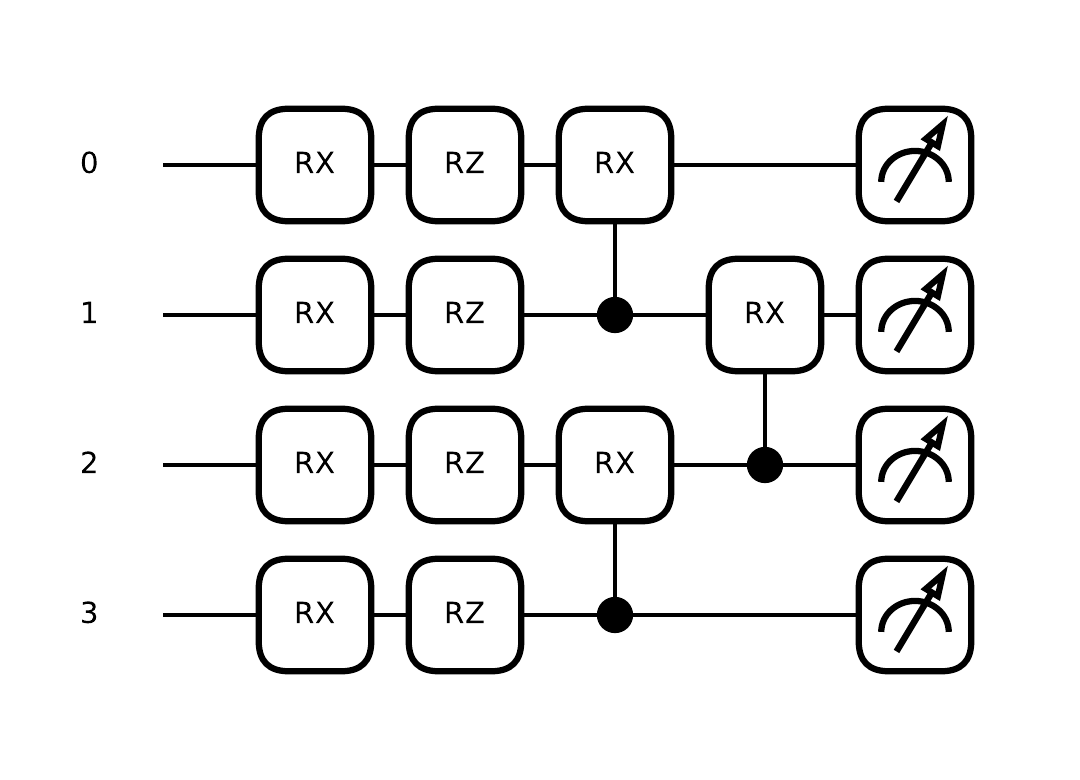}}
\hfill
\subfloat[Circuit 18]{%
  \includegraphics[width=0.45\columnwidth]{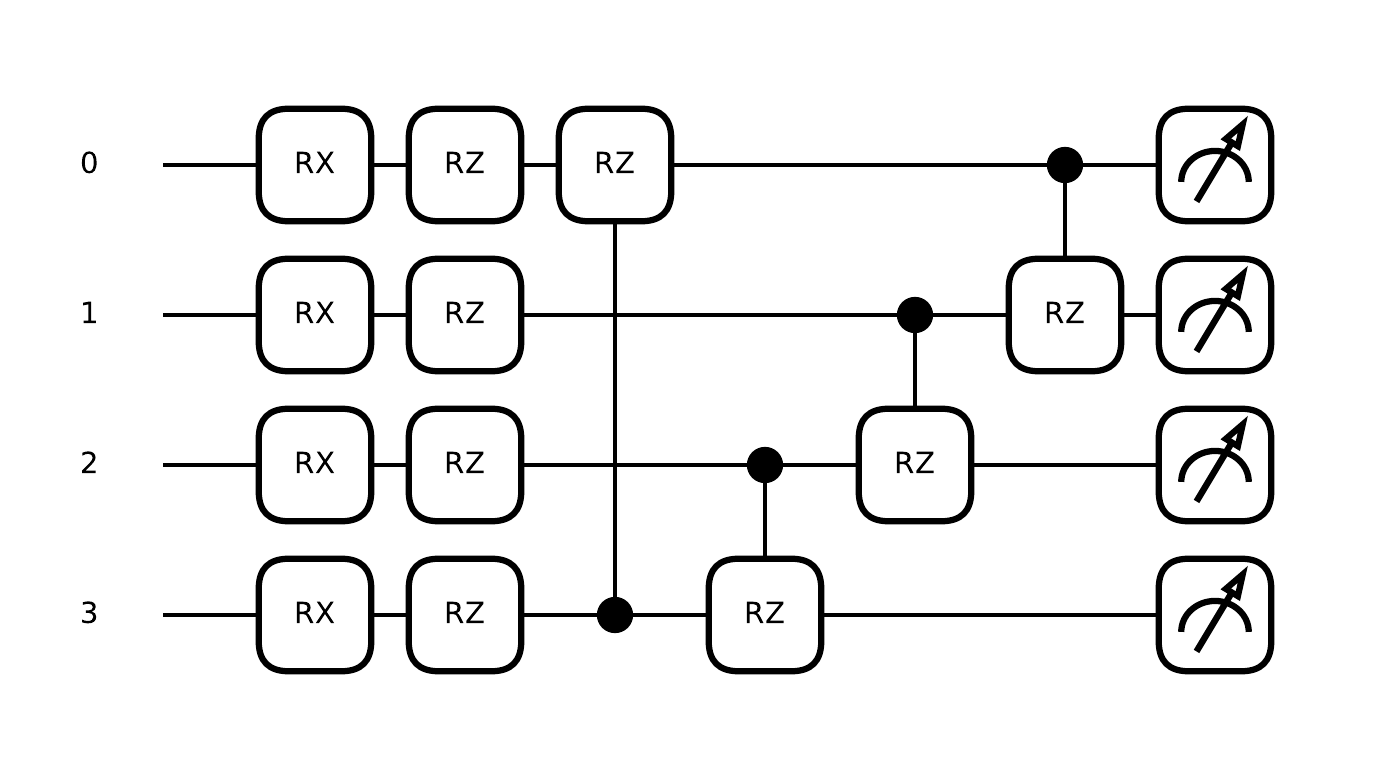}}

\vspace{0.3em}

\subfloat[Circuit 19]{%
  \includegraphics[width=0.45\columnwidth]{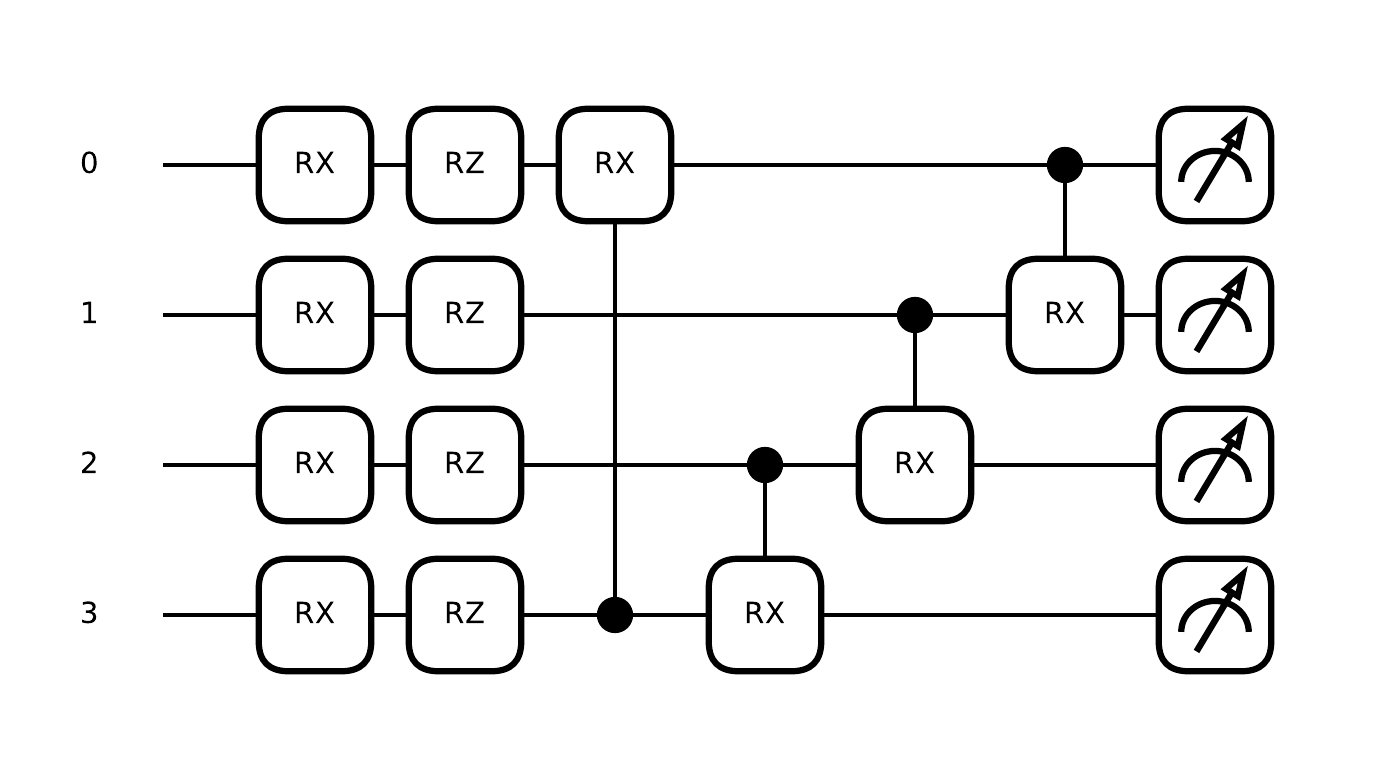}}
\hfill
\subfloat[NN Brickwork circuit]{%
  \includegraphics[width=0.45\columnwidth]{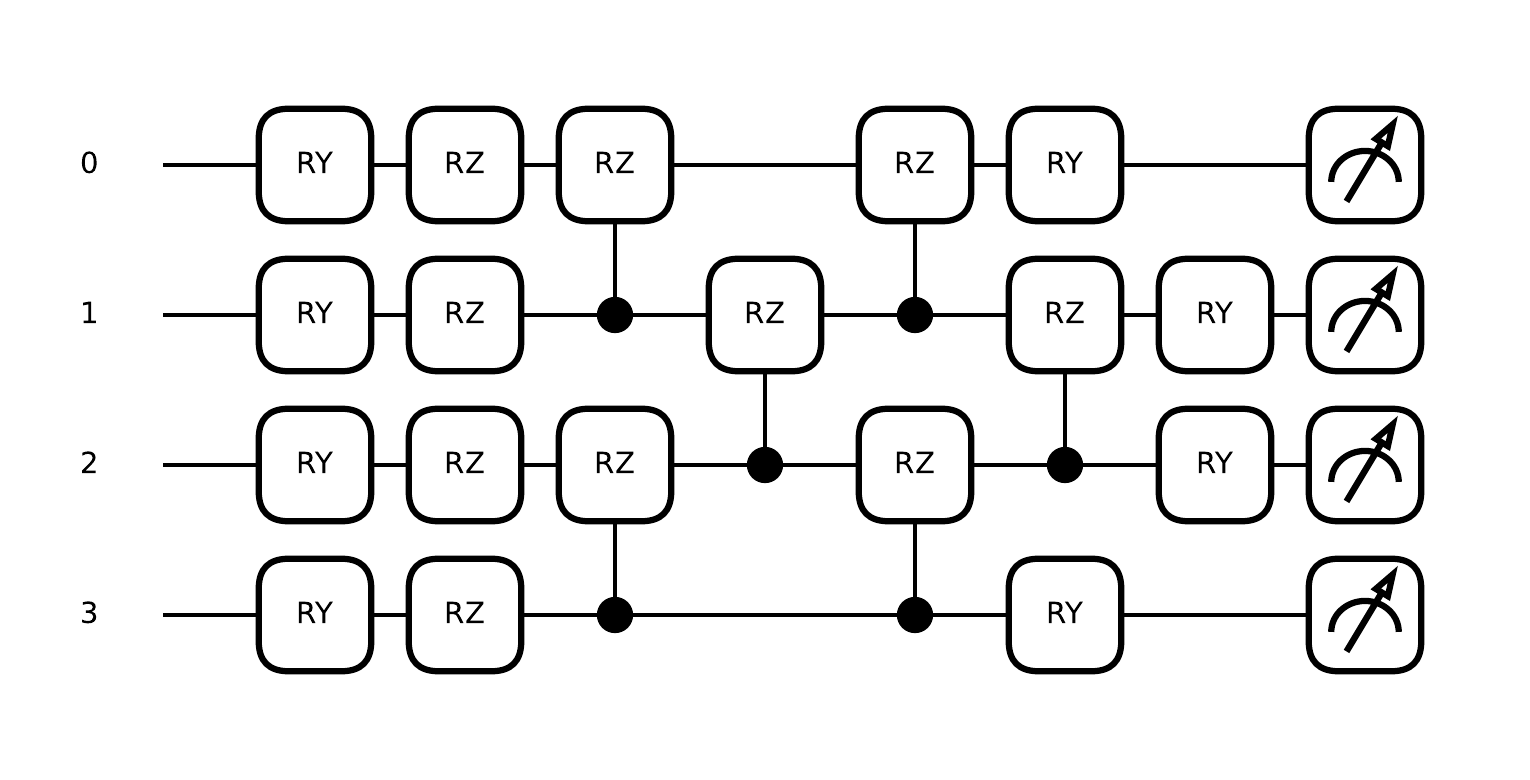}}

\vspace{0.3em}

\subfloat[NN Staircase circuit]{%
  \includegraphics[width=0.45\columnwidth]{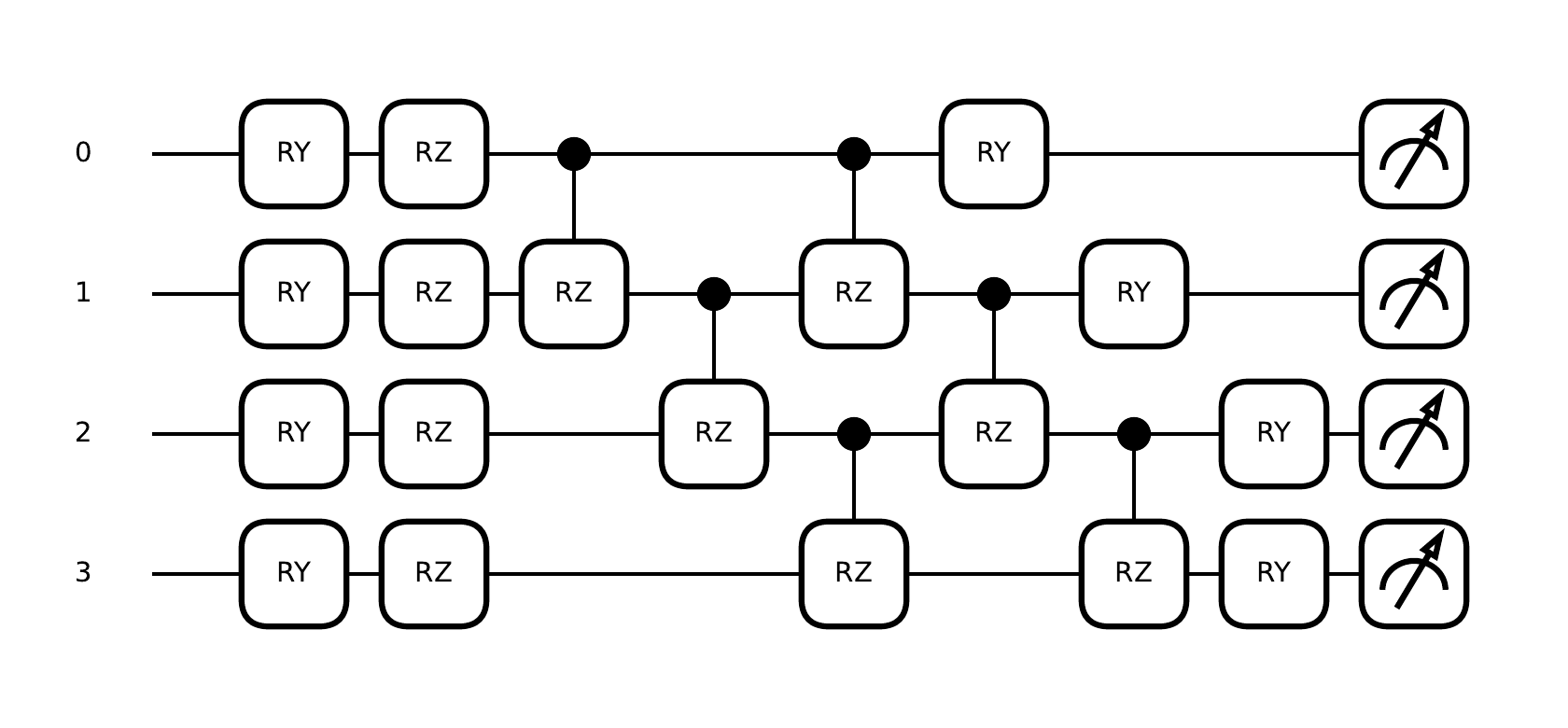}}
\hfill
\subfloat[PWR2 decreasing brickwork circuit]{%
  \includegraphics[width=0.45\columnwidth]{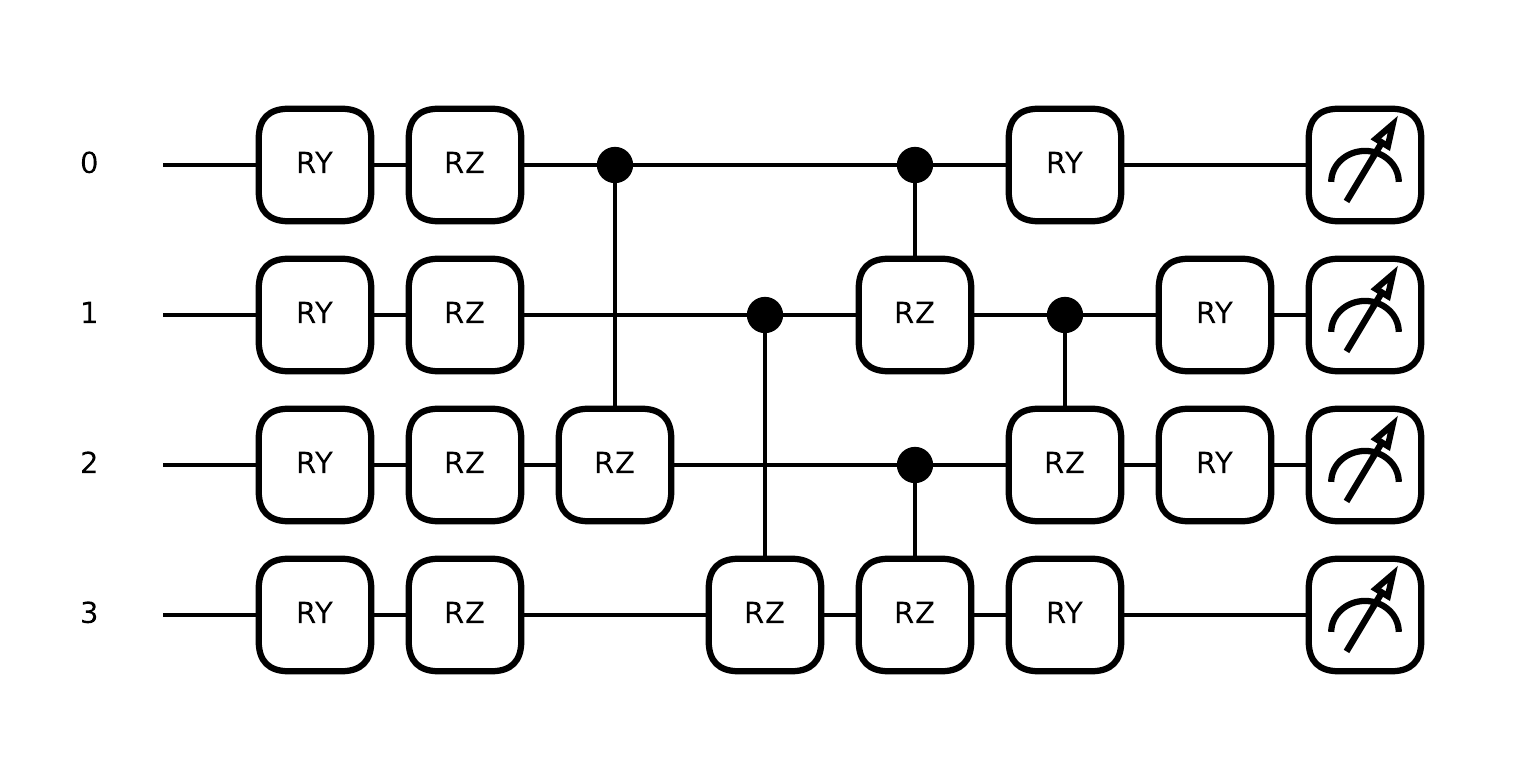}}

\vspace{0.3em}

\subfloat[PWR2 decreasing staircase circuit]{%
  \includegraphics[width=0.45\columnwidth]{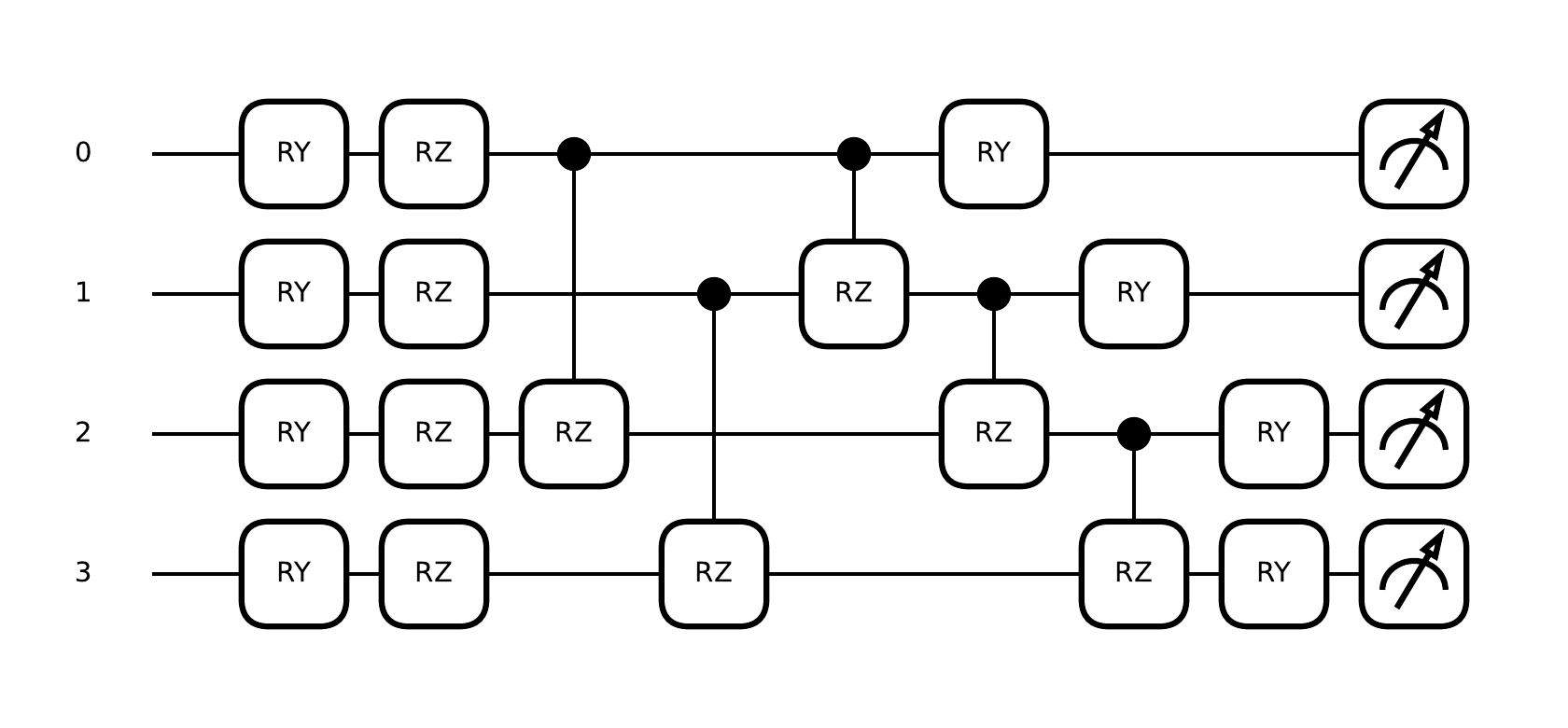}}
\hfill
\subfloat[PWR2 increasing brickwork circuit]{%
  \includegraphics[width=0.45\columnwidth]{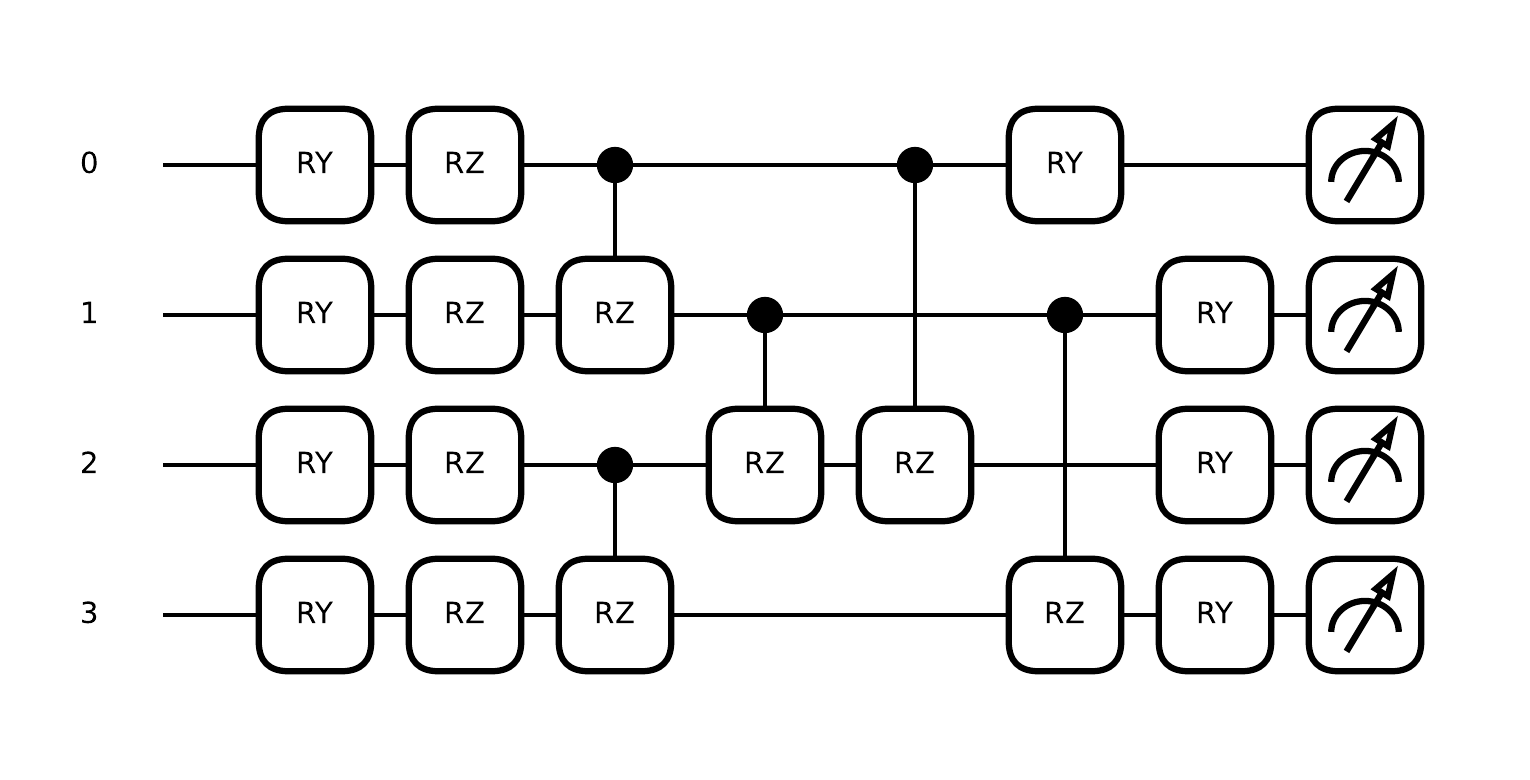}}

\vspace{0.3em}

\subfloat[PWR2 increasing staircase circuit]{%
  \includegraphics[width=0.45\columnwidth]{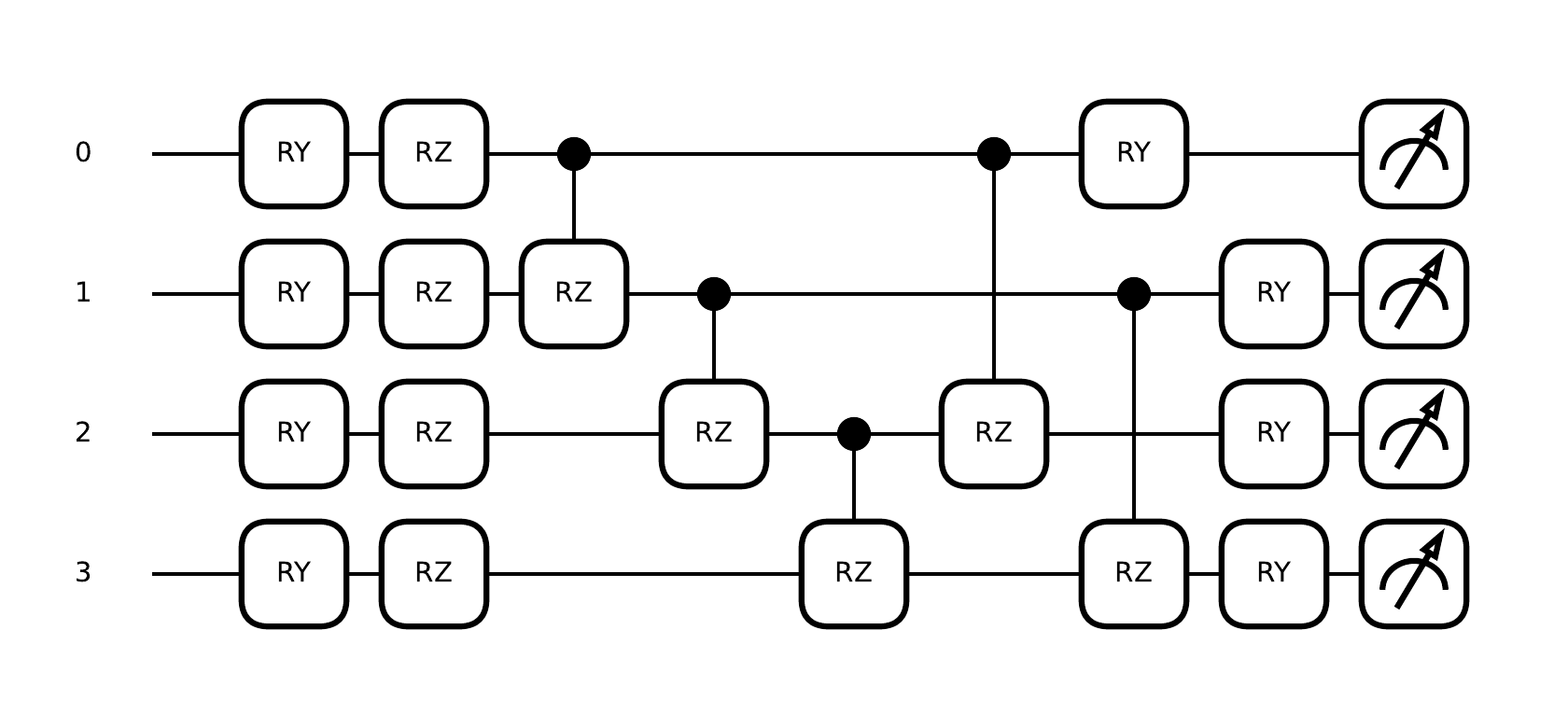}}

\caption{Ansätze used in the computation of
Fig.~\ref{fig:entangling_Cap_combined}(b). We depict one variational layer.}
\label{fig:ansatz}
\end{figure}

In Fig.~\ref{fig:ansatz} we show the circuits that were used to compute the expressibility and entangling capability shown in Fig.~\ref{fig:entangling_Cap_combined} in the main text.

\section{Additional VQE results}
\label{app:VQE_add}
\begin{figure}[H]
    \centering
    \includegraphics[width=\linewidth]{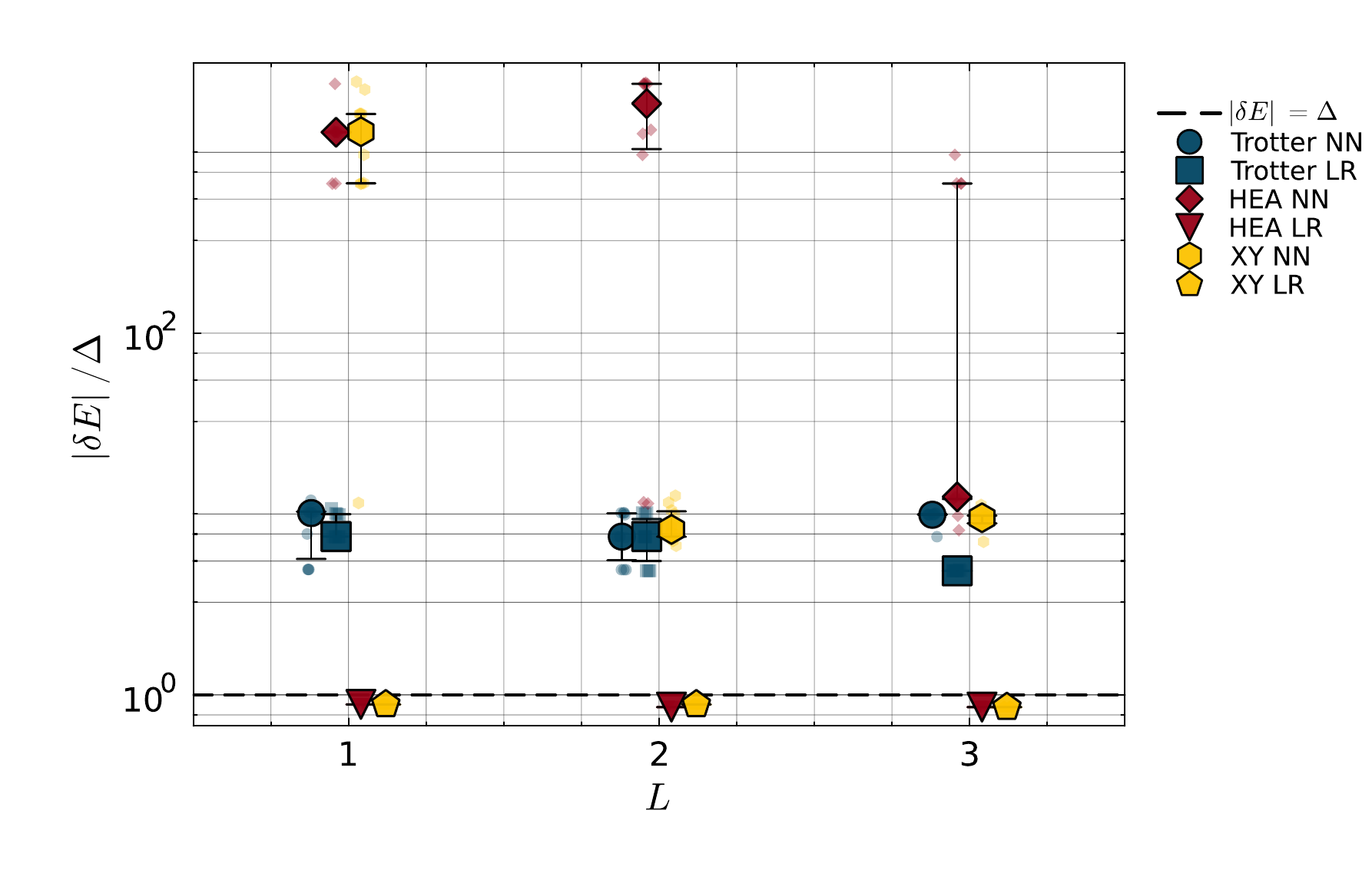}
    \caption{Energy error $|\delta E|/\Delta$ as a function of circuit depth $L$ for three ansatz families at $s = 5, B/J = 0.05$. Circles (filled) denote NN gates ($d = 1$) and squares (open) denote LR gates ($d = 8$). Each ansatz uses a fixed initialization across both gate ranges: $|0\rangle$ for the HVA, $|+\rangle$ for HEA, and XY. Large markers show the median over 10 random initializations; faint markers show individual seeds. The dashed line marks $|\delta E| = \Delta$. The NN–LR separation persists across all three ansatz families, indicating that it is a property of the gate connectivity rather than the ansatz structure.
}
    \label{fig:VQE_ansaetze}
\end{figure}

Here, we show additional results of the VQE in section \ref{sec:vqe_pwr2}. Fig.~\ref{fig:VQE_ansaetze} shows the robustness of our findings for different ansaetze. We compare the performance for three different ansaetze, the HVA from the main text, a HEA, and a RXY symmetry breaking ansatz. While the performance of the NN variants of the ansaetze differ, we find that for all three long-range ansaetze there is an initialization that reaches energy errors below the spectral gap. For consistency, we show the same initialization for the NN and LR variants of all three ansaetze. The chosen initialization corresponds to the best configuration found for the NN variants where the difference between initializations is much more pronounced than for the LR ansaetze.

\begin{figure}[H]
    \centering
    \includegraphics[width=\linewidth]{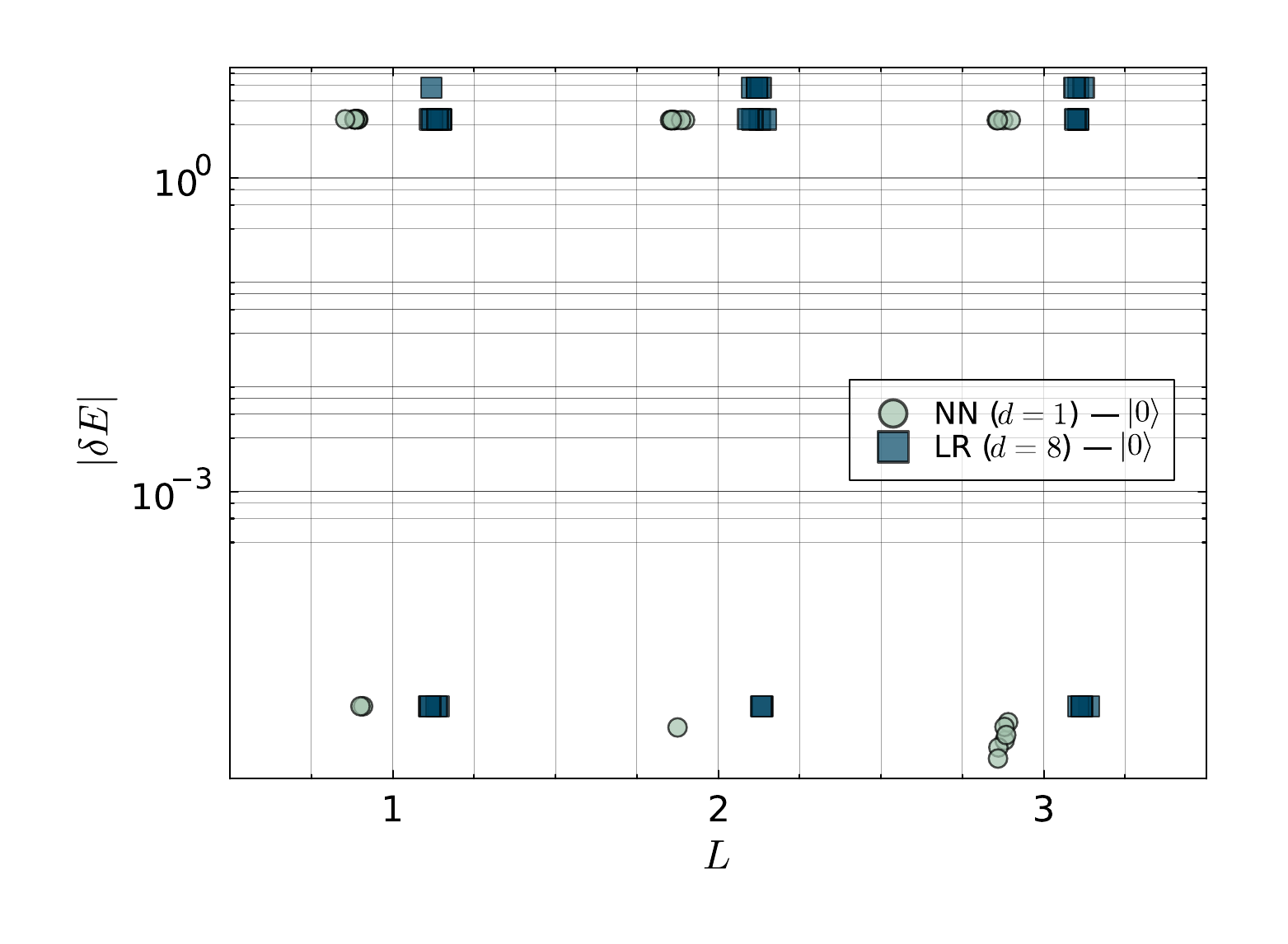}
    \caption{Absolute energy error $|\delta E|$ for direct VQE at $s = -5, B/J = 0.05$. Each dot represents one of 10 random initializations. The ground state is nearly doubly degenerate (gap $\sim 10^{-14}$), so we show $|\delta E|$ rather than $|\delta E|/\Delta$. Both NN and LR circuits with $|0\rangle$ initialization can reach the ground-state energy (bottom cluster), but the optimization landscape is fragmented into distinct basins. The NN circuit improves with depth, reaching the ground state in $\sim60\%$ of seeds at $L = 3$, whereas LR shows no clear depth dependence ($\sim30\%$). The $|+\rangle$ initialization (not shown) fails entirely for both gate ranges, with all seeds trapped at the initial-state energy $E= -Bn$.}
    \label{fig:VQE_s_Neg5}
\end{figure}

In Fig.~\ref{fig:VQE_s_Neg5} we show the results for VQE at $s=-5.0$, deep in the short-range interacting regime of the model. Here, we see that the disitribution is bimodal for both NN and LR gate ranges. However, while the success probability increases with depth for the NN ansatz we do not observe the analogous effect for the LR ansatz. 

\begin{figure}[H]
    \centering
    \includegraphics[width=\linewidth]{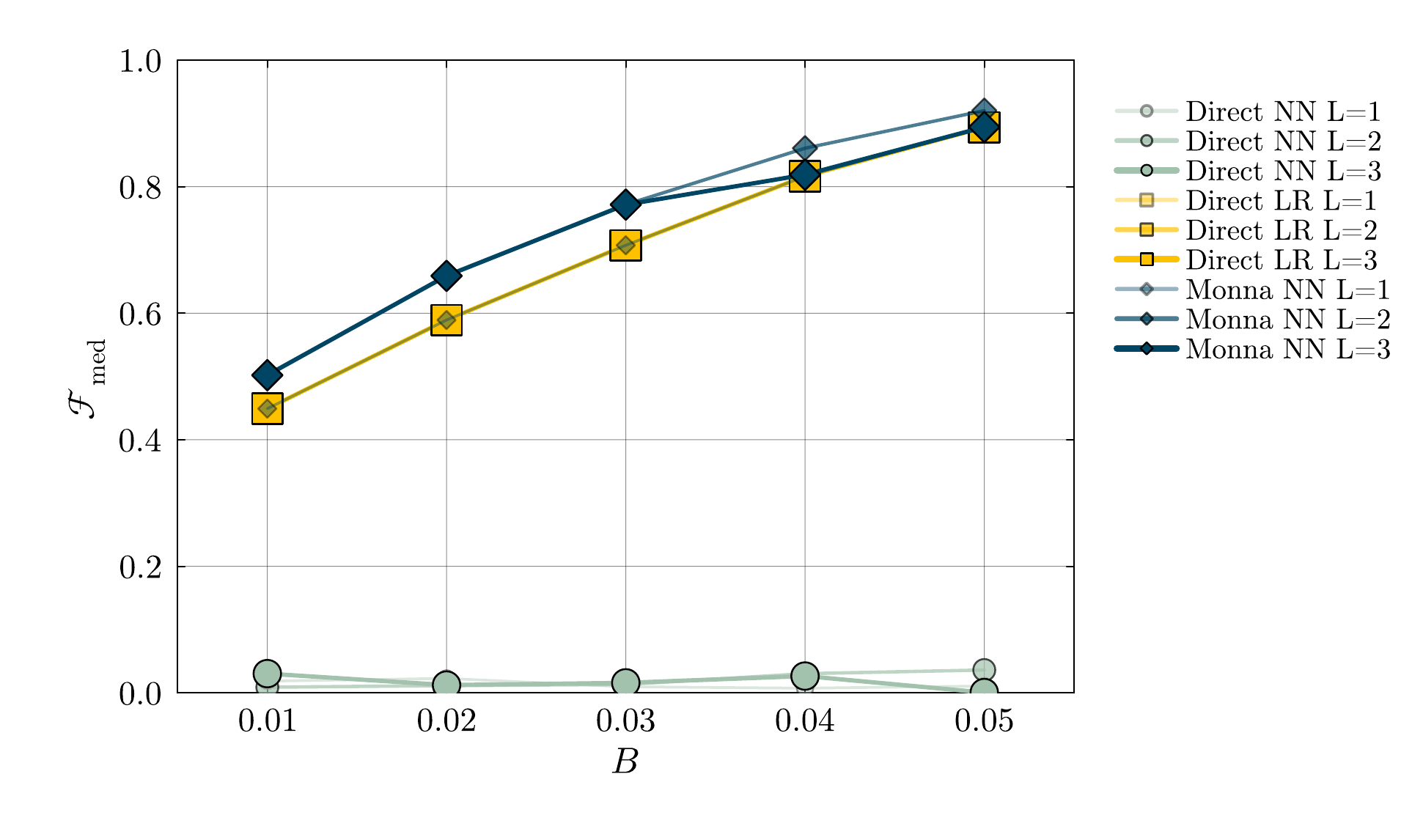}
    \caption{Median fidelity $\mathcal{F}_\mathrm{med}$ as a function of transverse field strength $B$ for $|+\rangle$ initialization at $s = 5$. Three curve families are shown at depths $L = 1, 2, 3$ (increasing opacity): Direct NN ($d = 1$), Direct LR ($d = 8$), and MonnaVQE NN ($d = 1$). Statistics are computed over 3 seeds per configuration. Fidelity increases with $B$ for both LR and Monna NN circuits but remains near zero for NN, confirming that the advantage of long-range connectivity and the Monna mapping persists across the range $B/J \in [0.01, 0.05]$}
    \label{fig:VQE_B_sweep}
\end{figure}

Finally, in Fig.~\ref{fig:VQE_B_sweep} we show the performance over a range of magnetic field values $B$ at $s=5.0$. We see that the median fidelity for the LR ansaetze and the included MonnaVQE ansaetze increases steadily with increasing the magnetic field while it stays near zero without a clear trend with increasing the magnetic field. We attribute the improvement in performance to the growing energy depth with increased magnetic field which the LR ansaetze and MonnaVQE can exploit to resolve the low-energy spectrum. The NN ansaetze, however, are not fundamentally limited by the spectral gap but by the failure to reach the low-energy entangled manifold. 

\section{Additional Monna map results}
\label{app:monna}
\begin{figure}[H]
    \centering
    \includegraphics[width=\linewidth]{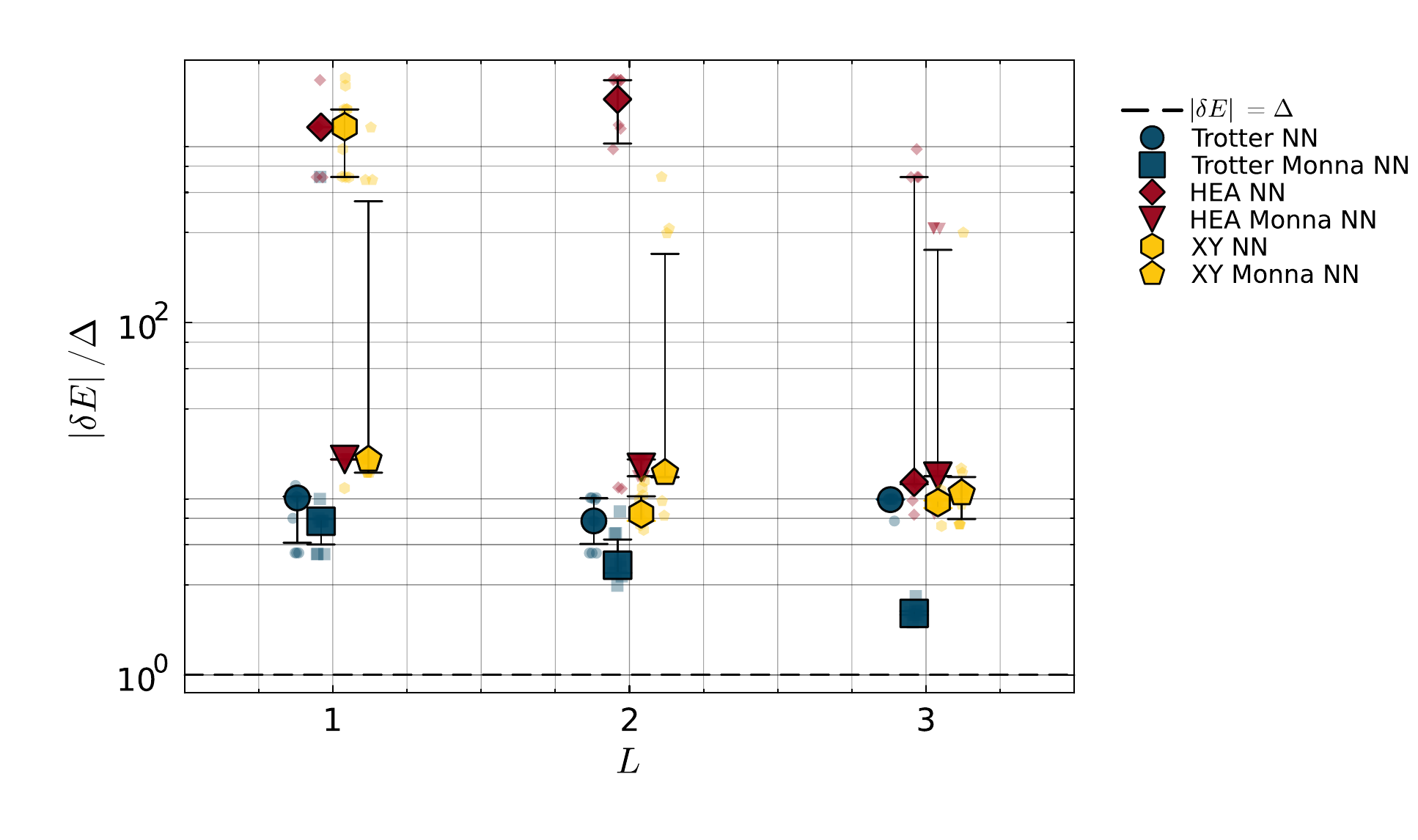}
    \caption{Energy error $|\delta E|/\Delta$ as a function of circuit depth $L$ comparing direct NN and Monna-mapped NN circuits across three ansatz families at $s = 5, B = 0.05$. Circles denote direct NN and diamonds denote Monna NN. Each ansatz uses a fixed initialization: $|0\rangle$ for HVA, $|+\rangle$ for HEA and XY. Large markers show the median over 10 random initializations; faint markers show individual seeds. The Monna mapping provides the largest improvement for the HVA (median $\sim2\Delta$ vs $\sim10\Delta$ for direct NN), while HEA and XY show smaller gains, consistent with those ansaetze also performing worse in the direct NN setting}
    \label{fig:Monna_ansaetze}
\end{figure}

\begin{figure}[H]
    \centering
    \includegraphics[width=\linewidth]{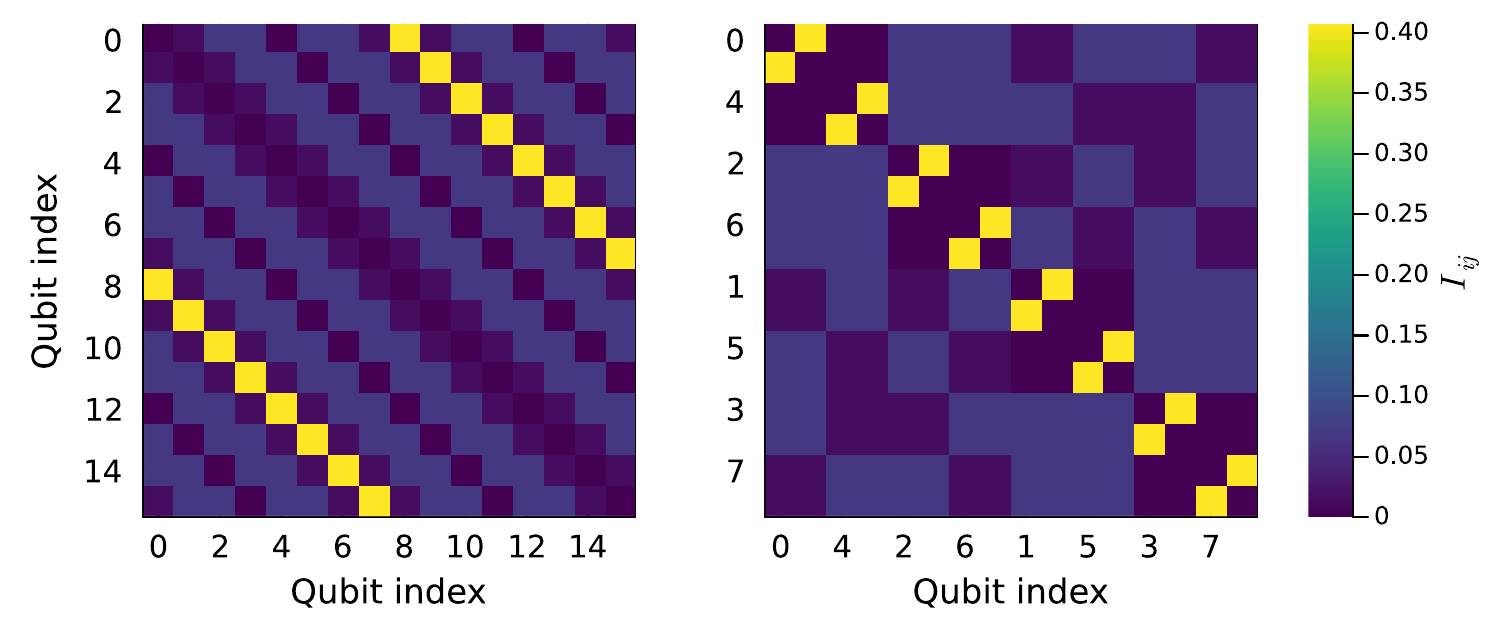}
    \caption{Mutual information of the ground state of the PWR2 model at $s=5.0$. The left panel shows the mutual information between all qubits in the original linear encoding. The right panel shows the mutual information of the ground state after applying the Monna map. It shows how the permutations can be used to localize the mutual information in the target state.}
    \label{fig:monn_MI}
\end{figure}

Here, we show for completeness how the VQE performance changes with different ansätze and initializations. \ref{fig:Monna_ansaetze} shows the results for direct VQE and MonnaVQE both using NN connectivity. We compare the HVA considered in the main text with a HEA and XY ansatz. We see that the big improvement we achieve with the MonnaVQE is smaller for the two different ansäetze. However, the other two ansaetze also perform worse on an absolute scale such that it might not be the long-range correlations that have limited the performance in the original VQE setting. Hence, the permutation is less effective in these scenarios.
In addition, we show that the analysis of the mutual information of the ground state of the PWR2 model confirms our Monna map approach. To this end, we compute the mutual information of the ground state both for the unpermuted and the permuted state. The mutual information is defined as 
\begin{equation}
    I_{ij}=S_i+S_j-S_{ij}
\end{equation}
where $S_A=-\mathrm{tr}\log \rho_A$ is the von-Neumann entropy of the reduced density matrix $\rho_A=\mathrm{tr}_B\rho^{AB}$. Fig.~\ref{fig:monn_MI} shows the results from numerically computing the mutual information for the exact ground states. It shows that the Monna map localizes the mutual information in the ground state. This aligns with our finding that the MonnaVQE performs better than the bare VQE in the main text.

In \cite{tkachenko_correlation-informed_2021} the MI is used in the cost function to find the permutation that achieves the most-localized mutual information map. In our case, the permutation is instead physically motivated and motivated on the level of the hamiltonian instead of the ground state.

\end{document}